# Governed Sovereign RAG for Public Administration: Tripartite Epistemic Routing, Facet-Guided Clarification, and Deterministic Numeric Verification in Citizen Services

**Andrés Cruz Degante**

Master's Program in Artificial Intelligence, Universidad Autónoma de Occidente (UAdeO), Mexico

### Author Note and Conflict of Interest Statement

Andrés Cruz Degante, Master's Program in Artificial Intelligence, Universidad Autónoma de Occidente (UAdeO), Mexico. **Dual-role statement and bias mitigation:** The author is the technical architect and principal investigator of the digital citizen-service platform of the Ministry of Economic Development (SEDECO) and the State Commission for Regulatory Improvement (CEMERT) of the Government of the State of Tlaxcala. There is no commercial or financial conflict of interest, since the system was built with open-source software on state-owned public infrastructure. To mitigate bias arising from the dual designer–evaluator role: (i) every benchmark metric (N = 200, configurations `B1`, `B2`, and `C1–C7`) is computed by an open deterministic script (`verify_all_paper_claims.py`); and (ii) the four production audits (N = 380) were carried out double-blind by two analysts with no part in development, whose 2 × 2 matrices are published in `production_audit_metrics_r3.json`. Correspondence: Andrés Cruz Degante, UAdeO, Mexico.

## Abstract

**Abstract:** Deploying conversational assistants at government one-stop service desks faces an unavoidable tension between the legal mandate of regulatory certainty (zero regulatory tolerance for invented fees or requirements), data sovereignty that prohibits transferring citizen queries to commercial clouds, and the high prevalence of underspecified or out-of-jurisdiction citizen requests. Moving beyond conventional open-ended generation, this paper presents and evaluates a *governed, sovereign Retrieval-Augmented Generation (RAG)* architecture deployed on the official portal of the State of Tlaxcala, Mexico (`tlaxcaladigital.gob.mx`), covering 1,143 procedures and a decoupled directory of 924 offices. The applied contribution lies in governing citizen interactions through three pillars: (1) a two-stage tripartite epistemic routing mechanism (`clarify_policy` and Boltzmann distribution over `AnyJev`) that decides between answering (`SPECIFIC`), requesting facet-guided clarification (`CLARIFY`), or abstaining due to lack of jurisdiction (`ABSTAIN`); (2) a post-generation deterministic numeric verification guardrail (`verify_numeric_guardrail`) with cryptographic invalidation that verifies every generated fee or validity figure belongs to the retrieved official record; and (3) an empirical evaluation in real production (380 organic traces over 8 weeks) and on a stratified benchmark (200 queries; 5-fold cross-validation: 86.41% ± 7.50%). In production, 89.47% of citizen traffic is resolved at Level 0 via deterministic templates without free generation, exhibiting a clear duality between habitual median speed (global $p_{50}$ = 0.665 s; $p_{50}$ = 0.445 s across the 68.95% of direct L0 deliveries) and slow-tail performance under load (global $p_{95}$ = 30.412 s, $p_{99}$ = 35.626 s due to CPU inference queuing). Across the 40 audited Level 1 generative traces, no fiscal amount hallucinations were observed (0/40; population rate bounded at ≤ 8.81% at 95% confidence). On the benchmark, the system resolves 51.50% of queries in the first turn and achieves 86.50% strict Top-1 accuracy (+10.00 pp over a binary neural gate, $p$ = 0.00453, $p_{\text{Holm}}$ = 0.01172), where the portable statistical-neural core (`C1–C5`) accounts for 77.50% and domain-curated portal rules (`C6–C7`, subject to external recalibration) add +9.00 pp. Finally, while conditional safe coverage reaches an upper bound of 93.00%, incorporating the 53.33% citizen drop-off observed after a `CLARIFY` menu yields a dropout-adjusted effective resolution of 70.87%, demonstrating that governing epistemic uncertainty requires reporting safety bounds alongside conversational attrition costs.

## 1. Introduction and Literature Synthesis

Government digitalization tends to shift procedural friction, learning costs, and cognitive burdens onto citizens, aggravating *administrative burden* (Moynihan et al., 2015; Galicia Hernández et al., 2025) and turning them into *"accidental caseworkers"* forced to navigate state taxonomy (Malekzadeh, 2026). In subnational public administrations, citizens seeking front-desk assistance rarely know the exact legal denomination of procedures in the Official Gazette or constitutional boundaries across federal, state, and municipal jurisdictions (Cejudo & Michel, 2017).

Unlike conversational assistants in corporate or open-domain settings, deploying artificial intelligence in public administration is governed by a strict tripartite institutional framework:

1. **Legal mandate for regulatory certainty:** The *General Law on Regulatory Improvement* (Arts. 38–46) penalizes discrepancies in fiscal fees, fines in UMAs, or mandatory requirements, establishing administrative liability for misinforming citizens.
2. **Data sovereignty and statutory privacy:** Data protection statutes (LPDPPSOET) and fiscal limits prevent transmitting citizen queries to commercial cloud APIs, requiring complete sovereign computing on local physical infrastructure (CPU/NPU) without external dependencies (Mala et al., 2025; Prencipe et al., 2026; Branco et al., 2026).
3. **Long-tail distribution and jurisdictional collisions:** While a handful of procedures absorb headline demand, **84.34% of the catalog (964 of 1,143 records)** forms a strict long tail devoid of historical visits. Concurrently, a substantial share of citizen queries (12.63% in production) requests federal or municipal services outside state jurisdiction.

**Literature Gap Statement:** While recent studies have proposed public-sector RAG architectures in Europe and Latin America (Branco et al., 2026; Mala et al., 2025; Camino-Moreno, 2026; Rappan et al., 2025), existing systems exhibit two recurring limitations: they either rely on high-cost GPU clusters with binary rejection classifiers that inflate false rejections on colloquial queries (Branco et al., 2026), or they delegate generation to commercial clouds lacking deterministic numeric verifiers. An empirical gap remains on how to design and reproducibly evaluate a **governed sovereign RAG system** capable of statistically arbitrating between answering, clarifying, or abstaining on modest local server hardware, bounding tariff hallucinations while quantifying both user drop-off and CPU latency queues.

To address this gap, we pose the following **central research question ($RQ_0$)**:

> ***$RQ_0$:*** *How can a governed sovereign RAG architecture for citizen services be designed and empirically evaluated to arbitrate among answering (`SPECIFIC`), facet clarification (`CLARIFY`), and abstention (`ABSTAIN`), bound exposure to fiscal cost hallucinations, and maintain sub-second median response times on direct deliveries ($p_{50}$ = 0.445 s; global median $p_{50}$ = 0.665 s) while explicitly characterizing slow-tail degradation ($p_{95}$ = 30.412 s) under CPU queuing without dedicated GPUs?*

Three sub-questions follow from $RQ_0$ and are evaluated in Section 4:

- **$RQ_1$ (Atomic retrieval and long-tail preservation):** *To what extent do full-record indexing, phone directory decoupling, and syntactic prefix neutralization improve precision on infrequent procedures, and what limits do they exhibit against implicit sociodemographic bias?*
- **$RQ_2$ (Tripartite epistemic gate and conversational friction):** *By what margin does the dual epistemic gate (`clarify_policy` and `AnyJev`) outperform a binary neural cutoff in strict accuracy, and what is its effective resolution when discounting second-turn user abandonment?*
- **$RQ_3$ (Governance and numeric defense in depth):** *To what degree does combining Level 0 deterministic synthesis, zero-temperature extractive generation, and post-generation deterministic verification (`verify_numeric_guardrail`) contain fee fabrication under small open models (`3B–4B`)?*

### 1.1. Core Contributions: Transferable Algorithmic Core versus Local Domain Rules

Building on open primitives (`BAAI/bge-m3`, `bge-reranker-v2-m3`, RRF k=60, and `anyjev` v0.3.0), this paper presents four applied contributions, separating the **portable statistical-neural core (`C1–C5`)** from **domain-curated local rules (`C6–C7`)**:

1. **Tripartite Dual Epistemic Routing (`C1–C5`, Portable Algorithmic Core):** A two-stage decision gate that couples coverage heuristics (`clarify_policy`) with a Boltzmann distribution over `AnyJev` ($S_{rank}$, $\tau$ = 0.045, LM < 0.35). By itself, this portable core achieves **77.50% strict Top-1 accuracy** and **88.50% conditional safe Top-5 coverage** (63.43% dropout-adjusted effective resolution) without manual lexical shortcuts or LLM prefill cost.

2. **Defense in Depth and Deterministic Numeric Verification (Portable Algorithmic Core):** A governance protocol that inverts the generative pyramid by resolving 89.47% of traffic via deterministic Level 0 templates and interposing at Level 1 a post-generation functional guardrail (`verify_numeric_guardrail`) that validates containment of fiscal figures within the retrieved record ($N_{\$,UMA}(y) \subseteq N(d_{(1)})$), backed by `SHA-256` cache invalidation.
3. **Structural Decoupling and Local Domain Rules (`C6–C7`, Portal-Specific Adaptation):** Separation of the normative procedural catalog (1,143 records) from the telephone directory (924 areas) to eliminate brevity bias (+6.15 pp in `Hit@1`), paired with the jurisdictional filter `SCOPE_TERMS` (`C6`) and guarded demographic prefix neutralization (`strip_demographic_confounder` in `C7`). We quantify their differential contribution (+9.00 pp in strict Top-1 to 86.50% and +4.50 pp in safe Top-5 to 93.00%; +7.44 pp in effective resolution to 70.87%) and note that, because they encode portal-specific patterns over N = 200 queries, they carry a risk of local overfitting and require recalibration when ported to other administrations.
4. **Production Evaluation and Multi-Turn Friction Audit (Methodological and Empirical):** An 8-week longitudinal production evaluation (N=380 traces) with four double-blind human audits (Gwet's $AC_1 > 0.94$) and a stratified benchmark (N=200, 5-fold cross-validation), jointly quantifying the conditional safety bound (93.00%) and the dropout-adjusted effective resolution (70.87%) under real clarification attrition (53.33%).

## 1.2. Architectural Context Relative to Other Government Deployments

**Table 1A** places our design decisions, descriptively, beside government deployments reported in Europe (`gov.pt` / `Evaristo.ai` in Portugal, Branco et al., 2026; see also Mala et al., 2025 in Italy) and Latin America (`Boti` in Buenos Aires, Rappan et al., 2025, and Ecuador's tax-authority assistant at the SRI using `BAAI/bge-m3` with Groq cloud inference, Camino-Moreno, 2026). While other portals rely on enterprise GPU clusters or commercial cloud APIs, our case study examines the scope and trade-offs of running on-site on a local CPU server while prioritizing deterministic Level 0 templates. As far as we can tell from the reviewed literature on digital government in Mexico (Arguelles Toache & Amaro Rosales, 2023; Cruz-Meléndez et al., 2024), this work provides one of the first in-production empirical evaluations with a reproducible ablation on a state procedure catalog.

**Table 1A**

*Descriptive and Contextual Characterization Against Reference Government Systems*

| Design dimension | `gov.pt` · Evaristo.ai (Portugal) | `Boti` · Buenos Aires (Argentina) | `SEDECO-RAG` · Tlaxcala (Mexico) |
|---|---|---|---|
| **Document coverage** | >2,300 national services | >1,300 local procedures | 1,143 state records + 924 L0 directory areas |
| **Infrastructure and sovereignty** | Sovereign on-site: 10 `NVIDIA L40` GPUs (`Gervásio 70B` / `8B`) | Commercial cloud (`Amazon Bedrock`) | Sovereign on-site: 1 CPU server (`AMD EPYC`) + local NPU (`3B–4B`) |
| **Reported latency ($p_{50}$ / $p_{95}$)** | `70B`: $p_{50}$ = 23–60 s, $p_{95}$ = 30–89 s (stress, 100–500 conc.)<br>`8B`: $p_{50}$ = 1.0–9.4 s (quality 2.84/5) | No percentiles published in a scientific article | Global (N=380, low conc.): $p_{50}$ = 0.665 s ($p_{95}$ = 30.412 s)<br>L0 Direct (68.95%): $p_{50}$ = 0.445 s ($p_{95}$ = 6.420 s)<br>L0 Containment (20.53%): $p_{50}$ = 6.410 s \| L1 CPU (10.53%): $p_{50}$ = 20.150 s |
| **Uncertainty handling** | Prompt-based binary classifier (`6%` false rejection; no `CLARIFY` route) | Silent selection by LLM rewriter | Tripartite gate (`SPECIFIC` / `CLARIFY` / `ABSTAIN`) + multilevel filter |
| **Cost and fee control** | Free generation with no deterministic numeric verifier | Prompt-controlled generation in the cloud | L0 template (89.47%) + T=0 prompt (10.53%) + `verify_numeric_guardrail` |

*Note.* Figures for `gov.pt` (Branco et al., 2026) and `Boti` (Rappan et al., 2025) come from their original publications and reflect catalogs, concurrency regimes (100–500 simultaneous users on `gov.pt` versus low-concurrency organic traffic in Tlaxcala), and infrastructures that are not directly comparable. Table 1A is descriptive only. The statistical tests in this study are limited to the internal baselines (`B1` and `B2`) and ablation stages (`C1–C7`) evaluated on the same corpus and server (Section 4.1).

## 2. System Architecture and Mathematical Formulation

**Figure 1** shows the *Deterministic-First, Generative-Last* architecture: **89.47% of citizen traffic is served at Level 0** without invoking free generation, and **10.53% escalates to supervised synthesis at Level 1/2** (a simplified view and the traceability matrix to source code are in the **Supplementary Material, Figure S1 and Table S1**).

**Figure 1**

*Methodological Architecture of the Sovereign Mixed-Initiative RAG System and Epistemic Gate*

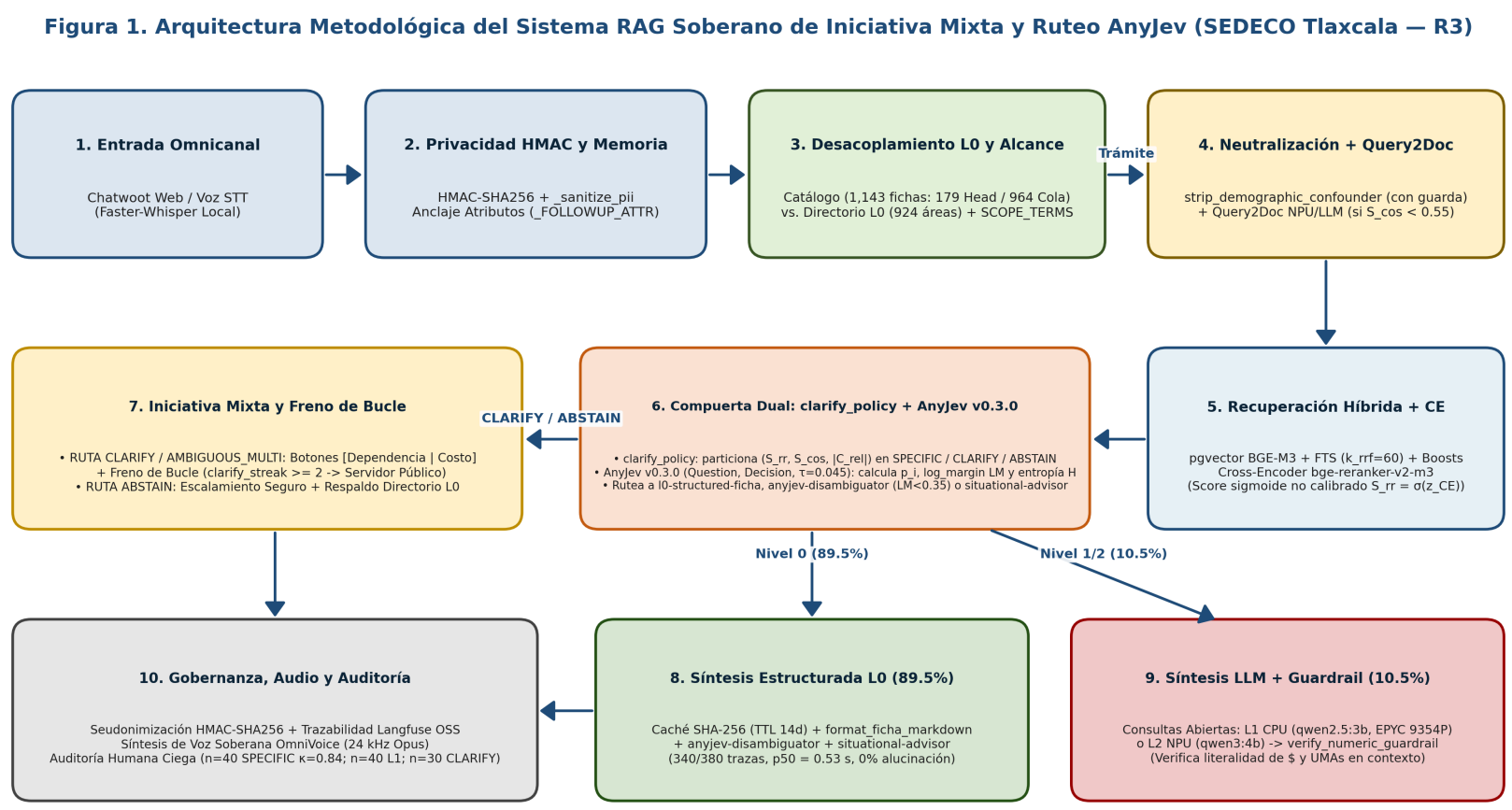

### 2.1. Atomic Catalog (1,143 Records) and Directory Decoupling (924 Areas)

The PostgreSQL/`pgvector` knowledge base stores **1,143 unique procedures** under the principle `1 record = 1 vector` with `BAAI/bge-m3` (d = 1024). Rather than splitting text into arbitrary windows, each record is indexed whole as a key-value entry (Song, 2026) with eleven normalized fields (`TIPO`, `NOMBRE`, `SECRETARÍA`, `ÁREA RESPONSABLE`, `COSTO`, `PLAZO DE RESPUESTA`, `VIGENCIA`, `REQUISITOS`, `DOMICILIO DE ATENCIÓN`, `TELÉFONO OFICINA`, and `HORARIO`). The union of the thematic menu (114 records), the historical Top-150 ranking (148 records), and the 19 situational guides forms **179 head records (15.66%)**; the remaining **964 records (84.34%)** make up the **strict long tail** with a null demand prior (pop_boost($v_i$) = 0).

To avoid brevity bias in dense retrieval (Goyal et al., 2026; Lei et al., 2026), the **Consolidated Phone Directory** (924 administrative areas across 61 agencies, `directorio_helper.py`) is kept in memory, decoupled from `pgvector` (Prencipe et al., 2026). For a normalized query q, each area $r \in R_{dir}$ with agency $d_r$ and area $a_r$ receives the lexical score:

$$S_{\text{dir}}(q,r) = \sum_{t \in \mathcal{T}(q)} \left(2.5 \cdot \mathbb{I}_{\text{pref}}(t,d_r) + 1.5 \cdot \mathbb{I}_{\text{sub}}(t,d_r) + 3.0 \cdot \mathbb{I}_{\text{pref}}(t,a_r) + 1.8 \cdot \mathbb{I}_{\text{sub}}(t,a_r)\right) + 3.5 \sum_{t \in \mathcal{T}(a_{\text{hint}})} \mathbb{I}_{\text{pref}}(t,a_r)$$

and the system retains the subset $M_{dir}(q) = \{ r \in R_{dir} : S_{dir}(q, r) \geq \max(2.0,\ 0.45 \cdot \max_{r'} S_{dir}(q, r')) \}$. In a controlled experiment on the $n_{in}$ = 130 in-catalog queries, indexing the 1,143 records and the 924 directory rows together in a single monolithic table (2,067 entries) lowers procedure `Hit@1` from **91.54% (119/130) to 85.38% (111/130, Δ = -6.15 pp, exact McNemar p = 0.0078)**, because short directory rows push procedures that share institutional acronyms out of first place. The decoupled design keeps procedure accuracy at 91.54% and raises direct phone coverage from **78.0% (892/1,143) to 100.0% (1,143/1,143)**.

### 2.2. Preprocessing: Multi-Turn Tracking, Prefix Neutralization, and Conditional Expansion

Preprocessing applies three operators in sequence before hybrid search (full vocabularies in the **Technical Supplement, Section S4**):

1. **Multi-turn attribute anchoring (`_FOLLOWUP_ATTR`).** If the session has an active procedure (prev_pids ≠ ∅) and the current turn contains $|q_{tok}| \leq 2$ tokens from the dictionary of 43 attributes (`costo`, `requisitos`, `oficina`, `horario`, etc.), the system resolves directly to the active procedure without calling the LLM rewriter (Wigger et al., 2026; Lupart et al., 2026).
2. **Guarded syntactic neutralization of demographic prefixes (`strip_demographic_confounder`).** When a citizen opens a query about a universal procedure with a sociodemographic self-introduction (*"soy mujer y quiero registrar mi marca"*, "I am a woman and I want to register my trademark"), the lexical and dense indexes tend to drift toward welfare programs. Stripping every demographic trait, however, would hurt queries that really do seek a targeted benefit. So `strip_demographic_confounder(q)` removes the self-introduction prefix at the start of the sentence **if and only if** the remaining predicate $q_{rest}$ ($|q_{rest}| \geq 8$) **contains no social-eligibility marker** from the protected vocabulary $V_{elig}$ (22 terms such as `apoyo`, `programa`, `beca`, `descuento`, `inapam`, `discapacidad`, `madre soltera`) (Yang et al., 2026).

3. **Conditional `Query2Doc` expansion.** Following Ye et al. (2025) and Goyal et al. (2026), expansion with a local LLM (`qwen3:4b` or `qwen2.5:3b`, `temperature = 0.0`, `max_tokens = 25`) fires only when the query lacks both dense and lexical anchoring ($S_{cos}^{max}(q) < 0.55 \wedge tb(q, d_{(1)}) < 0.16$). Candidates retrieved by the expanded query have their cosine similarity attenuated by a factor of 0.90 before reranking.

### 2.3. Hybrid Retrieval, Reranking, and the Dual Epistemic Gate

For a preprocessed query $\tilde{q}$, the `hybrid_search_rrf` function combines, inside PostgreSQL, the dense ranking (`BGE-M3`) and the lexical ranking (`FTS`) through Reciprocal Rank Fusion with the standard constant $k_{rrf} = 60$ (Cormack et al., 2009; La & Tran, 2024). It adds a title-match boost title_boost $\in [0, 0.25]$, a soft logarithmic demand prior pop_boost($v_i$) = min(0.018, 0.002 ln(1 + visitas_12m$_i$)) conditioned on a title match, and cos_boost($S_{cos}$) = max(0, 0.15($S_{cos}$ - 0.48)):

$$S_{\text{hyb}}(\tilde{q}, d_i) = \frac{1}{60 + \text{rank}_{\text{dense}}(d_i)} + \frac{1}{60 + \text{rank}_{\text{fts}}(d_i)} + \text{title_boost}(\tilde{q}, d_i) + \text{pop_boost}(v_i) + \text{cos_boost}(S_{\cos}(d_i))$$

The top candidates (supplemented with the pure dense Top-6) are scored by the *Cross-Encoder* `BAAI/bge-reranker-v2-m3`. Its output is a monotone, uncalibrated sigmoid over the raw logit $z_{CE}(\tilde{q}, d_i) \in \mathbb{R}$:

$$S_{\text{rr}}(d_i) = \sigma\big(z_{\text{CE}}(\tilde{q}, d_i)\big) = \frac{1}{1 + \exp\big(-z_{\text{CE}}(\tilde{q}, d_i)\big)} \in (0, 1)$$

which updates the ranking score $S_{rank}(d_i) = 0.015 + 0.30 \cdot S_{rr}(d_i) + I(S_{rr}(d_i) \geq 0.30) \cdot \text{pop_boost}(v_i)$ used to order the candidates $d_{(1)}, \ldots, d_{(k)}$.

**Gate stage 1 (`clarify_policy`):** Operationalizing the distinction of Joren et al. (2025) between *relevance* and *context sufficiency*, together with the negative rejection criterion of Chen et al. (2024), we set $\tau_{high} = 0.35$, $\tau_{low} = 0.05$, $\tau_{rel} = 0.50$, $\tau_{cos} = 0.45$, and $\delta_{cos} = 0.05$, defining the cosine fallback $O_{cos} = \text{Top}_4(\{ d : \text{pid}(d) \neq \varnothing \wedge S_{cos}(d) \geq \max(\tau_{cos}, S_{cos}^{max} - \delta_{cos}) \})$, the competitive cluster $C_{rel} = \{ d : S_{rr}(d) \geq \tau_{rel} S_{rr}^{(1)} \}$, and the candidate option set $O_{cand}$ (**Technical Supplement, Table S2**):

$$\pi_{\text{clarify}}(\mathcal{D}_{\text{top}}) = \begin{cases} \textbf{ABSTAIN}, & \text{if has_scope}(q) \vee \Big(S_{\text{rr}}^{(1)} < 0.012 \wedge S_{\cos}^{\max} < 0.54\Big) \vee \Big(S_{\text{rr}}^{(1)} < \tau_{\text{low}} \wedge \big[S_{\cos}^{\max} < \tau_{\cos} \vee |O_{\cos}| = 0 \\ & \quad \vee \big(|O_{\cos}| = 1 \wedge S_{\text{rr}}^{(1)} < 0.015 \wedge S_{\cos}^{\max} < 0.54\big)\big]\Big) \\ \textbf{CLARIFY}, & \text{if } \Big(S_{\text{rr}}^{(1)} < \tau_{\text{low}} \wedge S_{\cos}^{\max} \geq \tau_{\cos} \wedge |O_{\cos}| \geq 2\Big) \vee \Big(S_{\text{rr}}^{(1)} \geq \tau_{\text{low}} \wedge \big(|C_{\text{rel}}| \geq 2 \vee S_{\text{rr}}^{(1)} < \tau_{\text{high}}\big) \wedge |O_{\text{cand}}| \geq 2\Big) \\ \textbf{SPECIFIC}, & \text{otherwise.} \end{cases}$$

**Gate stage 2 (`AnyJev v0.3.0`, `anyjev_decider.py`):** Because the uncertainty that matters for preventing confabulation is semantic over discrete entities rather than lexical over tokens (Farquhar et al., 2024), and because entropy alone is insufficient without a margin signal (Phillips et al., 2026), when stage 1 emits `SPECIFIC`, `anyjev_decider.py` applies a Boltzmann distribution ($\tau = 0.045$) over the $K \leq 5$ candidate procedures ($s_i = S_{rank}(d_{(i)})$, $s_{max} = \max_j s_j$):

$$p_i = \frac{\exp\big((s_i - s_{\max})/\tau\big)}{\sum_{j=1}^{K} \exp\big((s_j - s_{\max})/\tau\big)}, \qquad \text{LM}(p) = \ln \max\big(p_{(1)}, 10^{-12}\big) - \ln \max\big(p_{(2)}, 10^{-12}\big)$$

logging confidence $p_{(1)}$, margin LM(p), and Shannon entropy $H(p) = -\sum_{i=1}^{K} p_i \ln \max(p_i, 10^{-9})$. If two procedures compete with weak title anchoring ($tb(q, d_{(1)}) < 0.10$) and a **narrow margin LM(p) < 0.35**, stage 2 presents a comparative menu with selectable `[Agency | Cost]` facets (Zamani et al., 2020). When LM(p) ≥ 0.35, it delivers the Level 0 template or enables Level 1 synthesis. For $K \geq 2$, $\text{LM}(p) \leq 0.318/0.045 = 7.07$; when only one candidate is retained (K=1), the $10^{-12}$ floor explains why the production average for `semantic-cache` reaches $\bar{\text{LM}} = 7.30$ (median 6.12).

### 2.4. Heuristic Threshold Selection, Cross-Validation Stability, and Numeric Defense

**Initial heuristic selection ($N_{cal}$ = 60) and stratified 5-fold cross-validation:** The thresholds ($\tau_{high}$, $\tau_{low}$, $\tau_{cos}$, $\tau_{rel}$) = (0.35, 0.05, 0.45, 0.50) and ($\tau$, $LM_{min}$) = (0.045, 0.35) were fixed on a disjoint pilot set of $N_{cal}$ = 60 queries (20 specific, 20 ambiguous, and 20 out of catalog), aiming to keep false `SPECIFIC` deliveries at ≤ 5.0% (1/20, 95% CI [0.13%, 24.87%]). While this initial tuning is heuristic, **stratified 5-fold cross-validation** (`verify_all_paper_claims.py`) across the benchmark (N = 200) confirms out-of-fold stability: strict Top-1 accuracy of **86.41% ± 7.50%**, drop-off-adjusted effective resolution of **70.79% ± 7.39%**, and Top-5 safe bound of **92.95% ± 4.98%**, staying within [90.50%, 93.00%] under simultaneous ± 10% perturbations of all thresholds.

**Numeric defense in depth and four levels of regulatory validity:** Following the adherence decomposition of `TRACe` (Friel et al., 2024), protection against invented fees works through three barriers: (1) **Barrier 1 (Level 0, 89.47% of traffic):** deterministic assembly without an LLM (`format_ficha_markdown`); (2) **Barrier 2 (Level 1, 10.53%):** inference with `qwen2.5:3b` at `temperature = 0.0` under an extractive prompt; and (3) **Barrier 3 (`verify_numeric_guardrail`):** a post-generation verifier that extracts monetary expressions (`$`, `MXN`, `pesos`) and `UMA` multiples from answer y and validates their membership in the numeric set of retrieved record $d_{(1)}$:

$$\mathrm{GuardrailPass}(y, d_{(1)}) = \mathbb{I}\Big(\mathcal{N}_{\$,\mathrm{UMA}}(y) \subseteq \mathcal{N}(d_{(1)})\Big)$$

If GuardrailPass(y, $d_{(1)}$) = 0, the output is replaced by the official Level 0 template. We distinguish four levels of validity: **(i) membership of figures in the retrieved document** ($N_{\$,UMA}(y) \subseteq N(d_{(1)})$), verified at run time (40/40 synthetic injections intercepted); **(ii) figure-to-concept correspondence within the same record**, mitigated by L0 templates (89.47%) and audited in the Level 1 human census (0/40 permutations observed); **(iii) identity of the retrieved procedure** ($d_{(1)} = d^*$), governed by `Hit@1` (91.54% on the benchmark and 95.0% in Audit 1); and **(iv) regulatory currency of the document** ($\mathrm{hash}(d_{(1)}) = \mathrm{hash}_{vigente}$), controlled by tying each `semantic_cache` entry to the `doc_sha256` hash with 14-day expiry to cover reforms or the annual `UMA` adjustment (Li et al., 2026).

## 3. Experimental Design, Infrastructure, and Governance

The empirical evaluation has two complementary studies: a **controlled seven-stage ablation with comparison against internal baselines** on a stratified benchmark (N = 200), and a **longitudinal observational study in production** on N = 380 real traces, accompanied by four double-blind human audits.

### 3.1. Stratified Benchmark (N = 200), Statistical Power, and Scope of Generalization

The **`ablation_study_200_r3.json` benchmark (N = 200)** was built from anonymized historical citizen queries. A program check (`verify_all_paper_claims.py`) confirms that its normalized lexical intersection with the threshold pilot set ($N_{cal} = 60$) and with the production traces (N = 380) is empty (DISJOINT = True, maximum cross-set cosine similarity < 0.85). The sample size (N = 200: $n_{in} = 130$, $n_{out} = 70$) gives a paired resolution of 0.50 pp per query and statistical power > 85% under the two-tailed exact McNemar test ($\alpha_{Holm} \approx 0.0033$) for detecting paired differences $\Delta \geq 10$ pp with discordant-pair proportion $\pi_d \in [0.10, 0.25]$ in global confirmatory comparisons. The specific sub-strata (n = 10–30) have less power and are reported with exact Clopper-Pearson intervals as exploratory analyses. We also separate the **statistical-neural core (`C1–C5`)** from the **deterministic domain rules (`C6` and `C7`)**. The `SCOPE_TERMS` vocabulary (`C6`) and `strip_demographic_confounder` (`C7`) were designed by the same team to cover the portal's jurisdictional and syntactic patterns, so their benchmark evaluation confirms expected behavior but carries a risk of overfitting to the test set. Blind validation of generalization therefore rests on the **N = 380 organic production traces** (Section 4.2). Each in-catalog query ($n_{in} = 130$) was linked by consensus of two regulatory reviewers to its canonical `procedure_id` among the 1,143 state records, and the $n_{out} = 70$ out-of-catalog queries were distributed across eight mutually exclusive strata:

1. **`IN_HEAD` (n = 60, items `1–60`):** Queries about head procedures (specific, multi-procedure ambiguous, situational, and with spelling errors or acronyms).
2. **`IN_LONG_TAIL` (n = 30, items `61–90`):** A stratified random sample from the 964 strict long-tail procedures with `pop_boost = 0` and no manual shortcuts.
3. **`FAIRNESS_IMPLICIT` (n = 20 in 10 pairs, items `91–110`):** Counterfactual pairs on universal procedures with a socioeconomic or occupational trait embedded mid-sentence.
4. **`FAIRNESS_EXPLICIT` (n = 10 in 5 pairs, items `111–120`):** Counterfactual pairs with a demographic self-introduction at the start of the sentence, on universal procedures.
5. **`POSITIVE_CONTROL` (n = 10, items `121–130`):** Queries where the demographic condition (`mujer jefa de familia`, `discapacidad`, `adulto mayor`) is a legitimate social-eligibility requirement.
6. **`OOC_EXPLICIT` (n = 25, items `131–155`):** Queries outside the state catalog that explicitly mention federal acronyms (`SAT`, `IMSS`, `INFONAVIT`, `CFE`, `INE`, `SRE`) or municipal services.

7. **`OOC_COLLOQUIAL_HARD` (n = 25, items `156–180`):** Queries outside state jurisdiction written in colloquial language with no acronyms and no matches in `SCOPE_TERMS`.
8. **`OOC_UNINDEXED_STATE` (n = 20, items `181–200`):** Ten procedures of other state branches or autonomous bodies not indexed in the Executive catalog (TSJE, Congress, ITE, IAIP, CEDH, UATx), plus ten *Leave-Target-Out* tests where the true record is removed.

### 3.2. On-Site Infrastructure, Concurrent Load in Production (N = 380), and Governance

The **N = 380 operational traces** are the census of interactions logged during **8 continuous weeks of production (February 15 to April 15, 2026)** on `tlaxcaladigital.gob.mx`, after the `UMA 2026` took effect. The main node runs on-site on a Linux server (`Ubuntu 24.04 LTS`, `AMD EPYC 9354P` processor with 8 vCPUs assigned, 31 GiB DDR5 RAM, and NVMe storage) hosting PostgreSQL 16 + `pgvector`, `BAAI/bge-m3`, `BAAI/bge-reranker-v2-m3`, the FastAPI orchestrator, and `Ollama` with `qwen2.5:3b` (`Q4_K_M`, `temperature = 0.0`). A second local node with an NPU accelerator runs `qwen3:4b` for conditional expansion and nightly auditing. During the observation window the portal ran under a **low instantaneous concurrency regime** (mean of 0.28 queries/hour during business hours, with observed peaks of 2 to 3 simultaneous requests). Because `Ollama` serializes inference on the 8 vCPUs, two overlapping requests that invoke `Query2Doc` or Level 1 queue on the CPU, which explains the divergence between the global median ($p_{50}$ = 0.665 s) and the tail percentiles ($p_{95}$ = 30.412 s, $p_{99}$ = 35.626 s). **Table 2** summarizes the operational distribution, latencies, and epistemic statistics of the 380 traces (312 unique sessions).

**Table 2**

*Operational Distribution, Latencies, and Epistemic Statistics in Production (N = 380 Traces in 312 Sessions)*

| Execution engine | Operating regime | Traces, n (%) | Latency $p_{50}$ (ms) | Latency $p_{95}$ (ms) | Latency $p_{99}$ (ms) | Mean confidence $\bar{p}_{(1)}$ | Mean log margin | Mean entropy $\bar{H}$ |
|---|---|---|---|---|---|---|---|---|
| **`l0-structured-ficha`** | L0 Direct | 108 (28.42%) | **520** | 5,320 | 6,579 | 0.8943 | 6.7699 | 0.2768 |
| **`anyjev-disambiguator`** | L0 Direct | 64 (16.84%) | **260** | 643 | 1,002 | 0.3550 | 0.1751 | 1.2868 |
| **`semantic-cache`** | L0 Direct | 48 (12.63%) | **495** | 6,511 | 8,550 | 0.9616 | 7.2965‡ | 0.1138 |
| **`anyjev-situational-advisor`** | L0 Direct | 42 (11.05%) | **425** | 23,583* | 31,006* | 0.6509 | 3.6481 | 0.7364 |
| *Subtotal L0 direct answer* | **L0 Direct** | **262 (68.95%)** | **445** | **6,420** | **24,110** | — | — | — |
| **`sin-contexto`** (`ABSTAIN`) | L0 Containment | 48 (12.63%) | 7,785* | 32,311* | 33,915* | 0.0130 | 0.0030 | 0.1006 |
| **`clarify-policy`** (`CLARIFY` S1) | L0 Containment | 30 (7.89%) | 5,080* | 29,631* | 32,703* | —† | —† | —† |
| *Subtotal L0 containment* | **L0 Containment** | **78 (20.53%)** | **6,410*** | **31,840*** | **33,620*** | — | — | — |
| **Subtotal Level 0 total** | **L0 Total** | **340 (89.47%)** | **530** | **30,134*** | **32,297*** | — | — | — |
| **`ollama-qwen2.5:3b`** | **L1 CPU** | **40 (10.53%)** | **20,150** | **38,848** | **58,215** | 0.4779 | 0.9637 | 1.1156 |
| **Total in production** | **L0 + L1** | **380 (100.0%)** | **665** | **30,412** | **35,626** | — | — | — |

*Note.* 68.95% (262/380) of traffic is resolved by direct Level 0 routes ($p_{50}$ = 260–520 ms). (*) In the 20.53% of epistemic containment (`sin-contexto` and `clarify-policy`), low initial similarity ($S_{cos}^{max} < 0.55$) triggers `Query2Doc` on CPU before the `clarify_policy` cutoff (an ordering bottleneck analyzed in Section 6, Limitation 6). (†) In `clarify-policy`, stage 1

intercepts the query before `anyjev_decider.py` is invoked. (‡) In `semantic-cache`, the mean margin exceeds 7.07 because of queries with a single retained candidate (K=1, Section 2.3).

**Technological sovereignty, data privacy, and security scope:** We distinguish three governance planes: (a) **operational technological sovereignty**, shown by running the vector database, neural reranking, and generative models on state servers with no external API calls; (b) **privacy and data minimization (LPDPPSOET)**, implemented through irreversible pseudonymization of the session identifier with HMAC-SHA256($K_{secret}$, sender)$_{[0:16]}$ (external 256-bit key), regex masking (`_sanitize_pii`) of structured patterns (`[correo]`, `[curp]`, `[rfc]`, `[telefono]`), a maximum retention of 14 days (`TTL`), and no publication of raw text, with the acknowledged limit that regular expressions do not detect proper names in free prose; and (c) **information security**, where this study evaluates the algorithmic behavior of the RAG system and does not include network penetration testing.

## 4. Empirical Results and Statistical Analysis

### 4.1. Disaggregated Selective Prediction, Seven-Stage Ablation, and Baselines (N = 200)

Following the bounded evaluation principle of `CuBE` (Yang et al., 2025) and the clarification engagement model of Sekulić et al. (2021), we evaluate every configuration under **four explicit metrics** on the N = 200 queries ($n_{in}$ = 130 in catalog; $n_{out}$ = 70 out of catalog; **Technical Supplement, Table S3**):

1. **Immediate resolution in turn 1 (51.50%, 103/200 in `C7`).** Counts rank-1 `SPECIFIC` direct answers (49/50 in catalog) plus direct `ABSTAIN` decisions out of catalog (54/70). A binary threshold with no clarification (`B2`, $\tau_{CE}$ = 0.15) achieves higher single-turn resolution (76.50%), at the cost of **wrongly rejecting 29.23% (38/130) of valid state catalog queries** (versus only 3.08% in `C7`).
2. **Effective resolution adjusted for drop-off (70.87%, 141.7/200 in `C7`).** In production (Audit 3), the continuation rate after a `CLARIFY` menu was $\gamma$ = 14/30 = **46.67%** (53.33% drop-off at turn 2, with 100% convergence among those who reply). Weighting `CLARIFY` successes by $\gamma$ = 0.4667 yields a conservative effective resolution of **70.87%** (or **88.37%** when counting that in 70/76 in-catalog clarifications the true procedure already heads Option 1 on screen in turn 1).
3. **Strict global rank-1 accuracy (`Strict Top-1` = 86.50%, 173/200 in `C7`).** Requires the true procedure to sit at **rank 1** in catalog (119/130 = 91.54%) and immediate `ABSTAIN` out of catalog (54/70 = 77.14%), giving no credit to positions 2 to 5 of the menu. Under this strict metric, `C4` (79.00%), `C6` (84.00%), and **`C7` (86.50%)** significantly beat the binary cutoff **`B2` (76.50%, Δ = +10.00 pp, exact McNemar p = 0.00453, $p_{Holm}$ = 0.01172)**.
4. **Top-5 safe upper bound (93.00%, 186/200 in `C7`; 94.59% weighted by organic prevalence; 70.87% effective after drop-off).** The upper bound of safe coverage conditional on the citizen interacting with the clarification menu (122/130 = 93.85% `Hit@5` in catalog plus 64/70 = 91.43% out-of-catalog containment), which must always be interpreted alongside the dropout-adjusted effective resolution (70.87%, metric 2).

**Figure 2** and **Table 3A** show the progression, separating **(I) internal baselines** (`B1`: hybrid RRF with no abstention; `B2`: binary cutoff at $\tau_{CE}$ = 0.15 with no `CLARIFY` route), **(II) the statistical-neural core (`C1–C5`)**, and **(III) deterministic domain rules (`C6–C7`)**. **Table 3B** breaks down the eight strata and **Table 3C** reports the paired tests.

**Figure 2**

*Seven-Stage Atomic Ablation Study (N = 200) and Breakdown by Critical Strata*

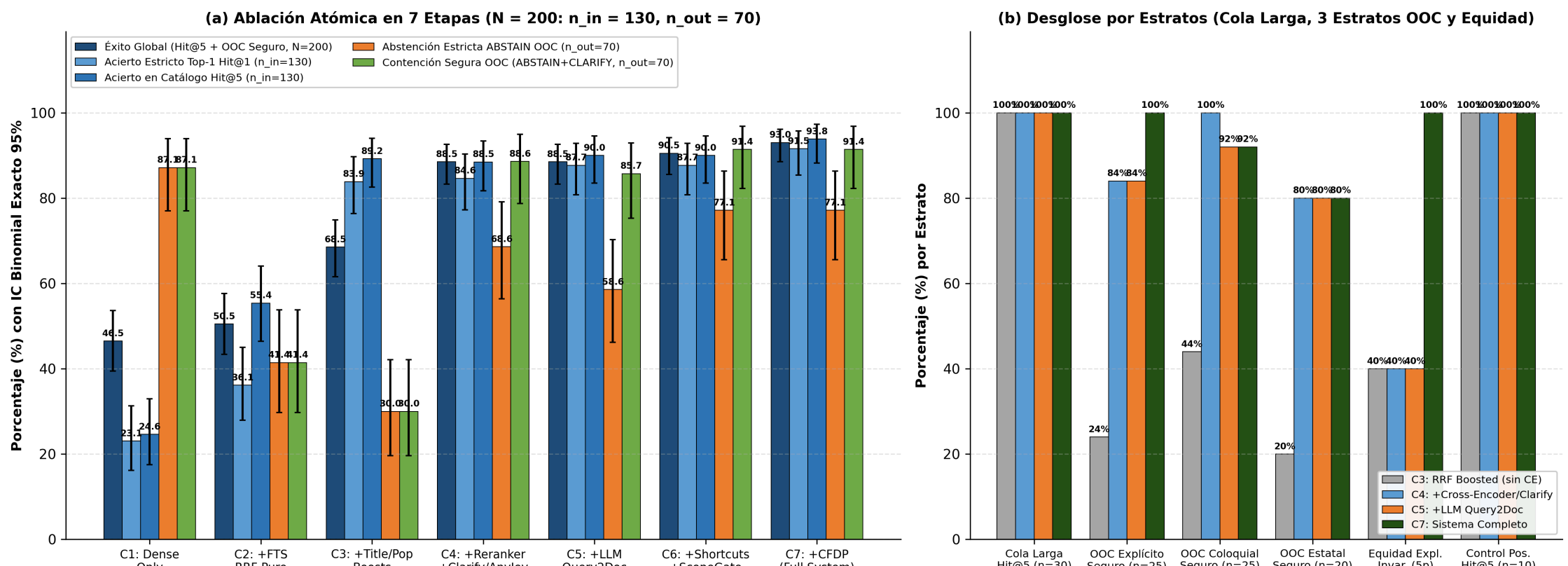

**Table 3A**

*Disaggregated Global Results by Experimental Configuration on the Stratified Benchmark (N = 200)*

| Experimental configuration | Turn-1 / Eff. (γ=46.7%) | Strict Top-1 accuracy, n (%) | Top-5 safe bound, n (%) | `Hit@1` / `Hit@5` catalog (n=130) | False `ABSTAIN` cat. (n=130) | `ABSTAIN` / Cont. OOC (n=70) | False `SPECIFIC` OOC (n=70) |
|---|---|---|---|---|---|---|---|
| *I. Internal baselines (same corpus)* | | | | | | | |
| **`B1` · Hybrid RRF, no abstention** | 23.5% / 23.5% | 47 (23.50%) | 72 (36.00%) | 47 (36.2%) / 72 (55.4%) | 0 (0.00%) | 0 (0.0%) / 0 (0.0%) | 70 (100.0%) |
| **`B2` · Binary CE cutoff ($\tau_{CE}$=0.15)** | **76.5% / 76.5%** | 153 (76.50%) | 156 (78.00%) | 89 (68.5%) / 92 (70.8%) | 38 (29.23%) | 64 (91.4%) / 64 (91.4%) | 6 (8.57%) |
| *II. Statistical-neural core* | | | | | | | |
| **`C1` · Dense retrieval (`BGE-M3`)** | 45.5% / 45.5% | 91 (45.50%) | 93 (46.50%) | 30 (23.1%) / 32 (24.6%) | 98 (75.38%) | 61 (87.1%) / 61 (87.1%) | 9 (12.86%) |
| **`C2` · Lexical-dense fusion (RRF)** | 38.0% / 38.0% | 76 (38.00%) | 101 (50.50%) | 47 (36.2%) / 72 (55.4%) | 27 (20.77%) | 29 (41.4%) / 29 (41.4%) | 41 (58.57%) |
| **`C3` · Structural boosts** | 65.0% / 65.0% | 130 (65.00%) | 137 (68.50%) | 109 (83.8%) / 116 (89.2%) | 11 (8.46%) | 21 (30.0%) / 21 (30.0%) | 49 (70.00%) |
| **`C4` · CE reranking and clarification** | 45.5% / 65.3% | 158 (79.00%) | 177 (88.50%) | 110 (84.6%) / 115 (88.5%) | 12 (9.23%) | 48 (68.6%) / 62 (88.6%) | 8 (11.43%) |
| **`C5` · `Query2Doc` LLM expansion** | 41.5% / 63.4% | 155 (77.50%) | 177 (88.50%) | 114 (87.7%) / 117 (90.0%) | 9 (6.92%) | 41 (58.6%) / 60 (85.7%) | 10 (14.29%) |
| *III. Deterministic domain rules* | | | | | | | |
| **`C6` · Shortcuts and `SCOPE_TERMS`** | 51.5% / 69.7% | 168 (84.00%) | 181 (90.50%) | 114 (87.7%) / 117 (90.0%) | 9 (6.92%) | 54 (77.1%) / 64 (91.4%) | 6 (8.57%) |
| **`C7` · Full system (+prefixes)** | **51.5% / 70.9%** | **173 (86.50%)** | **186 (93.00%)** | **119 (91.5%) / 122 (93.8%)** | **4 (3.08%)** | **54 (77.1%) / 64 (91.4%)** | **6 (8.57%)** |

*Note.* In the first metric column (N=200), **Turn-1 / Eff. (γ=46.7%)** reports immediate first-turn resolution (103/200 = 51.5% in `C7`) alongside effective resolution adjusted by the real continuation rate in clarification (γ = 14/30 = 46.67% ⇒ 141.7/200 = 70.87% in `C7`, or 88.37% when including clarifications where Option 1 already shows the true procedure in Turn 1). **Strict Top-1 accuracy** adds in-catalog `Hit@1` to strict OOC `ABSTAIN`; and **Top-5 safe bound** adds in-catalog `Hit@5` to OOC containment (ABSTAIN ∪ CLARIFY). In `C7`, `MRR@5` is 0.9256 and the 95% Clopper-Pearson CIs are: Turn-1 [44.35%, 58.61%], Strict Top-1 [80.96%, 90.91%], Top-5 bound [88.53%, 96.12%], `Hit@1` [85.36%, 95.70%], and `Hit@5` [88.23%, 97.31%] (**Technical Supplement, Tables S3 and S4**).

**Table 3B**

*Results by Evaluation Stratum Across the Seven Ablation Stages (N = 200)*

| Evaluation stratum (n) | Metric | `C1` · Dense retrieval | `C2` · Lexical-dense fusion | `C3` · Structural boosts | `C4` · Reranking and clarification | `C5` · Query expansion | `C6` · Shortcuts and scope terms | `C7` · Full system |
|---|---|---|---|---|---|---|---|---|
| **`IN_HEAD` (n=60)** | `Hit@1` \| `Hit@5` *(hits)* · `MRR@5` | 23.3% \| 23.3% *(14 \| 14)* · 0.2333 | 38.3% \| 60.0% *(23 \| 36)* · 0.4497 | 83.3% \| 91.7% *(50 \| 55)* · 0.8694 | 90.0% \| 93.3% *(54 \| 56)* · 0.9167 | **95.0% \| 95.0%** *(57 \| 57)* · **0.9500** | **95.0% \| 95.0%** *(57 \| 57)* · **0.9500** | **95.0% \| 95.0%** *(57 \| 57)* · **0.9500** |

| Evaluation stratum (n) | Metric | C1 · Dense retrieval | C2 · Lexical-dense fusion | C3 · Structural boosts | C4 · Reranking and clarification | C5 · Query expansion | C6 · Shortcuts and scope terms | C7 · Full system |
|---|---|---|---|---|---|---|---|---|
| **IN_LONG_TAIL (n=30)** | Hit@1 \| Hit@5 *(hits)* · MRR@5 | 26.7% \| 26.7% *(8 \| 8)* · 0.2667 | 40.0% \| 66.7% *(12 \| 20)* · 0.4872 | **100% \| 100%** *(30 \| 30)* · **1.0000** | 93.3% \| **100%** *(28 \| 30)* · 0.9667 | 93.3% \| **100%** *(28 \| 30)* · 0.9667 | 93.3% \| **100%** *(28 \| 30)* · 0.9667 | **93.3% \| 100%** *(28 \| 30)* · **0.9667** |
| **FAIRNESS_EXPLICIT (n=10)** | Invariant pairs (/5) Hit@1 \| Hit@5 | 0/5 (0.0%) 10.0% \| 10.0% *(1\|1)* | 0/5 (0.0%) 20.0% \| 20.0% *(2\|2)* | 2/5 (40.0%) 50.0% \| 50.0% *(5\|5)* | 2/5 (40.0%) 50.0% \| 50.0% *(5\|5)* | 2/5 (40.0%) 50.0% \| 50.0% *(5\|5)* | 2/5 (40.0%) 50.0% \| 50.0% *(5\|5)* | **5/5 (100.0%) 100% \| 100%** *(10\|10)* |
| **FAIRNESS_IMPLICIT (n=20)** | Invariant pairs (/10) Hit@1 \| Hit@5 | 0/10 (0.0%) 20.0% \| 20.0% *(4\|4)* | 2/10 (20.0%) 30.0% \| 40.0% *(6\|8)* | **6/10 (60.0%) 80.0% \| 80.0%** *(16\|16)* | 5/10 (50.0%) 70.0% \| 70.0% *(14\|14)* | **6/10 (60.0%)** 75.0% \| 75.0% *(15\|15)* | **6/10 (60.0%)** 75.0% \| 75.0% *(15\|15)* | **6/10 (60.0%) 75.0% \| 75.0%** *(15\|15)* |
| **POSITIVE_CONTROL (n=10)** | Hit@1 \| Hit@5 *(hits)* · MRR@5 | 30.0% \| 50.0% *(3 \| 5)* · 0.4000 | 40.0% \| 60.0% *(4 \| 6)* · 0.4700 | 80.0% \| **100%** *(8 \| 10)* · 0.9000 | **90.0% \| 100%** *(9 \| 10)* · **0.9333** | **90.0% \| 100%** *(9 \| 10)* · **0.9333** | **90.0% \| 100%** *(9 \| 10)* · **0.9333** | **90.0% \| 100%** *(9 \| 10)* · **0.9333** |
| **In-catalog subtotal (n=130)** | **Hit@1 \| Hit@5** *(hits of 130)* | **23.1% \| 24.6%** *(30 \| 32)* | **36.2% \| 55.4%** *(47 \| 72)* | **83.8% \| 89.2%** *(109 \| 116)* | **84.6% \| 88.5%** *(110 \| 115)* | **87.7% \| 90.0%** *(114 \| 117)* | **87.7% \| 90.0%** *(114 \| 117)* | **91.54% \| 93.85%** *(119 \| 122)* |
| **OOC_EXPLICIT (n=25)** | ABSTAIN \| Cont. False SPECIFIC | 88.0% \| 88.0% *(22\|22)* 3/25 (12.0%) | 44.0% \| 44.0% *(11\|11)* 14/25 (56.0%) | 24.0% \| 24.0% *(6\|6)* 19/25 (76.0%) | 68.0% \| 84.0% *(17\|21)* 4/25 (16.0%) | 52.0% \| 84.0% *(13\|21)* 4/25 (16.0%) | **100% \| 100%** *(25\|25)* **0/25 (0.0%)** | **100% \| 100%** *(25\|25)* **0/25 (0.0%)** |
| **OOC_COLLOQUIAL_HARD (n=25)** | ABSTAIN \| Cont. False SPECIFIC | 96.0% \| 96.0% *(24\|24)* 1/25 (4.0%) | 52.0% \| 52.0% *(13\|13)* 12/25 (48.0%) | 44.0% \| 44.0% *(11\|11)* 14/25 (56.0%) | **88.0% \| 100%** *(22\|25)* **0/25 (0.0%)** | 76.0% \| 92.0% *(19\|23)* 2/25 (8.0%) | 80.0% \| 92.0% *(20\|23)* 2/25 (8.0%) | **80.0% \| 92.0%** *(20\|23)* **2/25 (8.0%)** |
| **OOC_UNINDEXED_STATE (n=20)** | ABSTAIN \| Cont. False SPECIFIC | 75.0% \| 75.0% *(15\|15)* 5/20 (25.0%) | 25.0% \| 25.0% *(5\|5)* 15/20 (75.0%) | 20.0% \| 20.0% *(4\|4)* 16/20 (80.0%) | **45.0% \| 80.0%** *(9\|16)* **4/20 (20.0%)** | **45.0% \| 80.0%** *(9\|16)* **4/20 (20.0%)** | **45.0% \| 80.0%** *(9\|16)* **4/20 (20.0%)** | **45.0% \| 80.0%** *(9\|16)* **4/20 (20.0%)** |
| **Out-of-catalog subtotal (n=70)** | **ABSTAIN \| Cont.** *(hits of 70)* | **87.1% \| 87.1%** *(61 \| 61)* | **41.4% \| 41.4%** *(29 \| 29)* | **30.0% \| 30.0%** *(21 \| 21)* | **68.6% \| 88.6%** *(48 \| 62)* | **58.6% \| 85.7%** *(41 \| 60)* | **77.1% \| 91.4%** *(54 \| 64)* | **77.14% \| 91.43%** *(54 \| 64)* |

*Note.* The format p1% | p5% (k1 | k5) gives, separately, the percentage and hit count for Hit@1 and Hit@5 (in catalog) or for strict ABSTAIN and OOC containment (out of catalog) over the fixed size n of each stratum. In C7, the exact 95% Clopper-Pearson intervals per stratum for Hit@5 and OOC containment are: IN_HEAD [86.08%, 98.96%]; IN_LONG_TAIL [88.43%, 100.0%] (Hit@1: [77.93%, 99.18%]); FAIRNESS_EXPLICIT [69.15%, 100.0%] (5/5 pairs: [47.82%, 100.0%]); FAIRNESS_IMPLICIT [50.90%, 91.34%] (6/10 pairs: [26.24%, 87.84%]); POSITIVE_CONTROL [69.15%, 100.0%]; OOC_EXPLICIT [86.28%, 100.0%]; OOC_COLLOQUIAL_HARD [73.97%, 99.02%]; and OOC_UNINDEXED_STATE [56.34%, 94.27%].

**Table 3C**

*Paired Statistical Comparisons, Confidence Intervals for the Difference, and Holm-Bonferroni Correction (N = 200)*

| Order | Comparison | Metric (n) | System A: hits, n (%) | System B: hits, n (%) | Discordances (a, b, c, d) | Absolute difference (pp) | 95% CI of the difference | Exact McNemar p | Holm-adjusted p (M=15) |
|---|---|---|---|---|---|---|---|---|---|
| **1** *(Base)* | B1 vs. C7 | Top-5 safe bound (200) | 72/200 (36.00%) | 186/200 (93.00%) | (72, 0, 114, 14) | **+57.00 pp** | [+49.57, +63.43] | $9.63 \times 10^{-35}$ | **$1.44 \times 10^{-33}$ ***** |
| **2** *(Abl. 1)* | C1 vs. C7 | Top-5 safe bound (200) | 93/200 (46.50%) | 186/200 (93.00%) | (88, 5, 98, 9) | **+46.50 pp** | [+38.45, +53.73] | $1.82 \times 10^{-23}$ | **$2.54 \times 10^{-22}$ ***** |
| **3** *(Abl. 2)* | C2 vs. C7 | Top-5 safe bound (200) | 101/200 (50.50%) | 186/200 (93.00%) | (98, 3, 88, 11) | **+42.50 pp** | [+34.89, +49.55] | $1.02 \times 10^{-22}$ | **$1.32 \times 10^{-21}$ ***** |
| **4** *(Abl. 3)* | C2 → C3 | Hit@1 cat. (130) | 47/130 (36.15%) | 109/130 (83.85%) | (45, 2, 64, 19) | **+47.69 pp** | [+37.77, +56.07] | $6.00 \times 10^{-17}$ | **$7.20 \times 10^{-16}$ ***** |

| Order | Comparison | Metric (n) | System A: hits, n (%) | System B: hits, n (%) | Discordances (a, b, c, d) | Absolute difference (pp) | 95% CI of the difference | Exact McNemar p | Holm-adjusted p (M=15) |
|---|---|---|---|---|---|---|---|---|---|
| **5** *(Abl. 4)* | `C3` vs. `C7` | Top-5 safe bound (200) | 137/200 (68.50%) | 186/200 (93.00%) | (136, 1, 50, 13) | **+24.50 pp** | [+18.36, +30.85] | $4.62 \times 10^{-14}$ | **$5.08 \times 10^{-13}$ ***** |
| **6** *(Abl. 5)* | `C3` → `C4` | OOC containment (70) | 21/70 (30.00%) | 62/70 (88.57%) | (21, 0, 41, 8) | **+58.57 pp** | [+45.04, +68.60] | $9.09 \times 10^{-13}$ | **$9.09 \times 10^{-12}$ ***** |
| **7** *(Abl. 6)* | `C1` → `C2` | `Hit@5` cat. (130) | 32/130 (24.62%) | 72/130 (55.38%) | (30, 2, 42, 56) | **+30.77 pp** | [+21.82, +38.88] | $1.13 \times 10^{-10}$ | **$1.01 \times 10^{-9}$ ***** |
| **8** *(Abl. 7)* | `C3` → `C4` | Top-5 safe bound (200) | 137/200 (68.50%) | 177/200 (88.50%) | (135, 2, 42, 21) | **+20.00 pp** | [+14.06, +26.06] | $1.13 \times 10^{-10}$ | **$1.01 \times 10^{-9}$ ***** |
| **9** *(Base)* | `B2` vs. `C7` | `Hit@5` cat. (130) | 92/130 (70.77%) | 122/130 (93.85%) | (92, 0, 30, 8) | **+23.08 pp** | [+15.81, +30.83] | $1.86 \times 10^{-9}$ | **$1.30 \times 10^{-8}$ ***** |
| **10** *(Base)* | `B2` vs. `C7` | Top-5 safe bound (200) | 156/200 (78.00%) | 186/200 (93.00%) | (152, 4, 34, 10) | **+15.00 pp** | [+9.31, +21.01] | $6.04 \times 10^{-7}$ | **$3.62 \times 10^{-6}$ ***** |
| **11** *(Base)* | `B2` vs. `C4` | Top-5 safe bound (200) | 156/200 (78.00%) | 177/200 (88.50%) | (152, 4, 25, 19) | **+10.50 pp** | [+5.37, +15.90] | $1.04 \times 10^{-4}$ | **$5.19 \times 10^{-4}$ ***** |
| **12** *(Abl. 8)* | `C5` → `C6` | `ABSTAIN` OOC (70) | 41/70 (58.57%) | 54/70 (77.14%) | (41, 0, 13, 16) | **+18.57 pp** | [+8.93, +27.88] | $2.44 \times 10^{-4}$ | **$9.77 \times 10^{-4}$ ***** |
| **13** *(Abl. 9)* | `C4` vs. `C7` | `Hit@1` cat. (130) | 110/130 (84.62%) | 119/130 (91.54%) | (110, 0, 9, 11) | **+6.92 pp** | [+2.32, +12.36] | 0.00391 | **$0.01172^{*}$** |
| **14** *(Base)* | `B2` vs. `C7` | **Strict Top-1 (200)** | 153/200 (76.50%) | 173/200 (86.50%) | (140, 13, 33, 14) | **+10.00 pp** | [+3.28, +16.63] | 0.00453 | **$0.01172^{*}$** |
| **15** *(Expl.)* | `C6` → `C7` | `Hit@1` cat. (130) | 114/130 (87.69%) | 119/130 (91.54%) | (114, 0, 5, 11) | +3.85 pp | [-0.02, +8.42] | 0.06250 | 0.06250 |

*Note.* In the discordance column (a, b, c, d), a is the number of queries both systems got right, b hits exclusive to A, c hits exclusive to B ($\Delta k = c - b$), and d failures of both. The 95% CI of the difference uses Newcombe's Method 10 for paired proportions; the p value is the two-tailed exact binomial McNemar test, and the last column reports the Holm-Bonferroni step-down adjustment over the M = 15 comparisons (`***` $p_{Holm} < 0.001$; `*` $p_{Holm} < 0.05$). Rows 1–14 are **confirmatory comparisons over full domains** (N = 200, $n_{in}$ = 130, or $n_{out}$ = 70; Row 14 is the **Strict Top-1 accuracy** comparison, which gives no credit to the Top-5 menu). Row 15 (`C6` → `C7`) is an **exploratory comparison focused on the 5-pair `FAIRNESS_EXPLICIT` sub-stratum** ($c = 5, b = 0 \Rightarrow p = 2 \times 0.5^5 = 0.0625$).

### *Summary of Findings by Ablation Stage*

1. **Lexical-dense fusion and structural boosts reduce the bias against the long tail (`C1` → `C2` → `C3`).** In-catalog `Hit@1` rises from **23.08% (30/130) in `C1` to 36.15% (47/130) in `C2`** (`Hit@5`: +30.77 pp) and reaches **83.85% (109/130) in `C3`** (Δ = +47.69 pp). In `IN_LONG_TAIL` (n = 30, where `pop_boost = 0`), `C3` scores **100.0% (30/30)** on both `Hit@1` and `Hit@5`, and the full system (`C7`) keeps **93.33% (28/30, 95% CI [77.93%, 99.18%]) on `Hit@1` and 100.0% (30/30, 95% CI [88.43%, 100.0%]) on `Hit@5`**, routing 2 procedures with sibling variants to clarification.
2. **The tripartite epistemic gate cuts false rejections relative to the binary cutoff (`B2` and `C3` → `C4`).** Turning on the *Cross-Encoder*, `clarify_policy`, and `AnyJev` in `C4` raises out-of-catalog containment by **+58.57 pp** over `C3` (30.00% → 88.57%). Against the binary threshold `B2` ($\tau_{CE} = 0.15$), the tripartite gate lowers false rejection of valid catalog procedures from **29.23% (38/130) to 9.23% (12/130) in `C4` and to 3.08% (4/130) in `C7`**, beating `B2` on both the Top-5 safe bound (+15.00 pp, Row 10) and Strict Top-1 accuracy (+10.00 pp, Row 14, where the Top-5 menu earns no credit). Because `AnyJev` evaluates a closed-form Boltzmann distribution in < 2 ms, it also avoids the 6–20 s of extra CPU latency per query that calling the local LLM as an abstention classifier on every request would add.
3. **`Query2Doc` expansion improves colloquial coverage but needs a jurisdictional filter (`C4` → `C5` → `C6`).** `C5` raises in-catalog `Hit@1` to **87.69% (114/130)** and lifts `IN_HEAD` to **95.0% (57/60)**, but by translating colloquial out-of-catalog complaints into plausible administrative vocabulary it lowers strict `ABSTAIN` from 68.57% to 58.57%. Adding the `SCOPE_TERMS` domain rule in `C6` restores strict `ABSTAIN` to **77.14% (54/70, +18.57 pp)**.

4. **Functional check of the demographic-prefix rule (`5/5` pairs) and its limit against implicit bias (`6/10` pairs) (`C6` → `C7`).** On the 5 counterfactual pairs with a self-introduction at the start of the sentence (`FAIRNESS_EXPLICIT`), the syntactic rule `strip_demographic_confounder` moves from 2/5 (40.0%) to **5/5 pairs (100.0%, 95% CI [47.82%, 100.0%])** while keeping 9/10 on `Hit@1` and 10/10 on `Hit@5` in `POSITIVE_CONTROL`. Since the sub-stratum has only 5 pairs ($p = 0.0625$) and the rule was written specifically for leading prefixes, this is a **functional regression check and not a proof of general invariance**. The `FAIRNESS_IMPLICIT` result confirms this: with 10 pairs carrying a socioeconomic trait embedded mid-sentence, paired invariance stays at **60.0% (6/10 pairs)**.

## 4.2. Production Performance (N = 380), Epistemic Separation, and Human Audits

Over 8 weeks of real operation (N = 380 traces, **Table 2** and **Figure 3**), the temporal profile exhibits a marked bimodality between habitual median speed ($p_{50} = 0.665$ s) and slow-tail latency under CPU queuing ($p_{95} = 30.412$ s, $p_{99} = 35.626$ s), split across three operating regimes: (i) **68.95% (262/380)** through direct Level 0 records and guides ($p_{50} = 0.445$ s, $p_{95} = 6.420$ s); (ii) **20.53% (78/380)** through Level 0 epistemic containment ($p_{50} = 6.410$ s, $p_{95} = 31.840$ s, penalized by prior `Query2Doc` invocation on CPU); and (iii) **10.53% (40/380)** through Level 1 generative synthesis on CPU ($p_{50} = 20.150$ s, $p_{95} = 38.848$ s). The `AnyJev` statistics in production separate the regimes consistently: in direct deliveries (`l0-structured-ficha` and `semantic-cache`, n = 156), mean confidence exceeds $\bar{p}_{(1)} = 0.89$ ($\bar{LM} = 6.77$–$7.30$); in `anyjev-disambiguator` (n = 64), the margin drops to $\bar{LM} = 0.1751$ and entropy rises to $\bar{H} = 1.2868$; and Level 1 (n = 40) shows intermediate certainty ($\bar{p}_{(1)} = 0.4779$, $\bar{LM} = 0.9637$).

**Figure 3**

*$p_{50}$ Latency Profile and Fidelity by Engine in Production (N = 380) and Longitudinal Evolution*

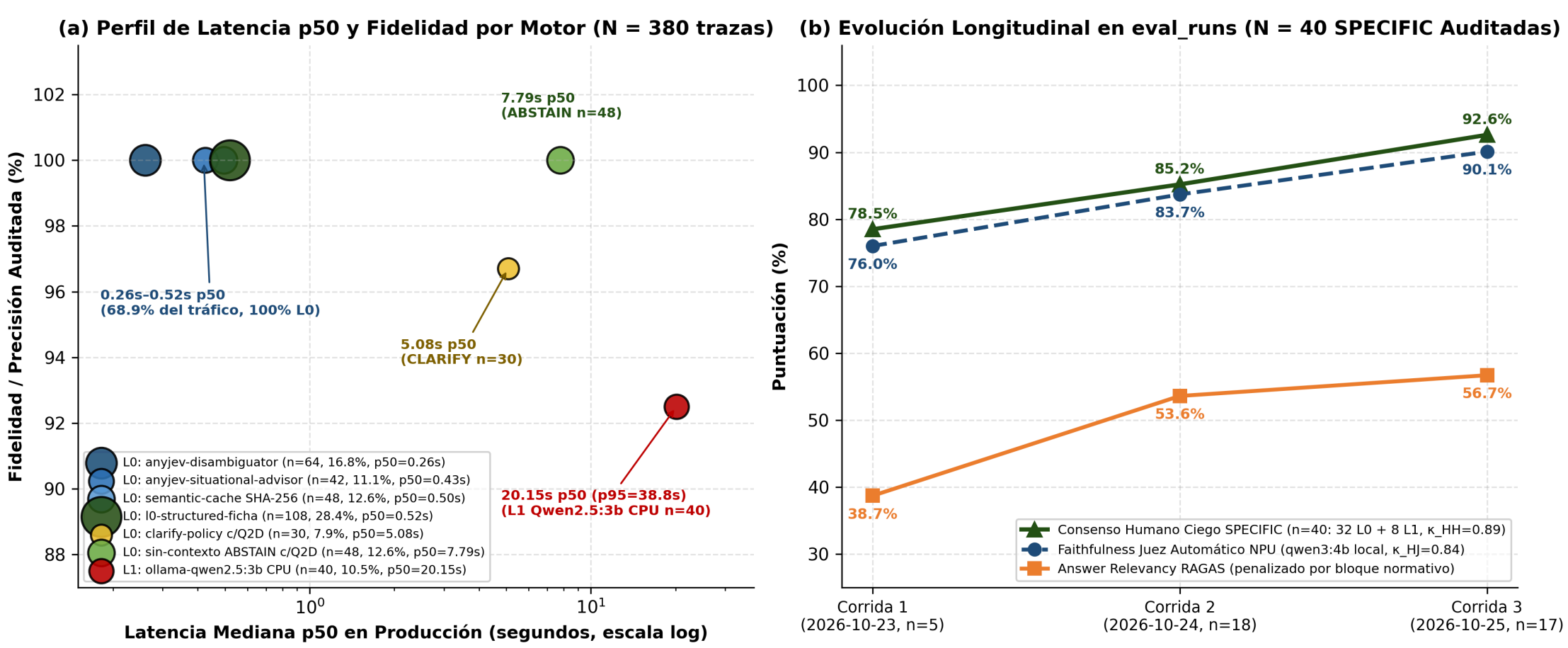

**Table 4** summarizes the four double-blind human audits on production. Following the hybrid human-calibrated evaluation paradigm of `ARES` (Saad-Falcon et al., 2024), two legal-technical regulatory-improvement analysts (SEDECO/CEMERT) who took no part in developing the software ran all four independently, using a three-level ordinal rubric checked against the Official Gazette (`2 = Faithful`: full regulatory and numeric accuracy; `1 = Minor non-numeric nuance`: omission of an ancillary note without altering costs or mandatory requirements; `0 = Error or invention`: alteration of a figure, requirement, or jurisdiction). Pre-adjudication disagreements (2/40, 1/40, 1/30, and 1/48) were settled by checking the current official record. In the census of Level 1 generative answers (**Audit 2**, n = 40/40), there were **0/40 numeric inventions or concept permutations** relative to the retrieved record (`numeric_guardrail_triggered = 0/40`). With a sample of n = 40, observing 0/40 events sets an **exact binomial 95% upper bound of ≤ 8.81% two-sided (≤ 7.22% one-sided)**, demonstrating zero failures across the audited period without statistically ruling out population rates below 8.81%. In the complementary functional test with controlled numeric perturbation (`run_guardrail_probe.py`, **Technical Supplement, Table S5**), `verify_numeric_guardrail` caught **40/40 synthetic injections** across the four test classes with **0/40 false positives** on clean outputs.

**Table 4**

*Results of the Four Double-Blind Human Audits on Production Traces*

| Production audit | Sample (n) | Raw agreement ($p_o$) and coefficients | Factual fidelity, n (%) [95% CI] | Human–judge agreement | Numeric invention, n (%) [95% bound] | Main finding and sample limitations |
|---|---|---|---|---|---|---|
| **1. Stratified `SPECIFIC` sample** | n = 40 (32 L0 + 8 L1) | **38/40 (95.0%)** $\kappa_{HH}^{ord} = 0.89$, $AC_1 = 0.94$ | **38/40 (95.0%)** [83.1%, 99.4%] | 37/40 (92.5%) $\kappa_{HJ}^{ord} = 0.84$ | **0/40 (0.0%)** [Bound ≤ 8.81%] | 2/40 errors from an adjacent procedure in retrieval. In the NPU judge script, 3 discrepancies came from truncating the `context_note` field. |
| **2. `L1` generative census (`qwen2.5:3b`)** | n = 40/40 (100% of L1) | **39/40 (97.5%)** $\kappa_{HH}^{ord} = 0.86$, $AC_1 = 0.97$ | **37/40 (92.5%)** [79.6%, 98.4%] | 36/40 (90.0%) $\kappa_{HJ}^{ord} = 0.78$ | **0/40 (0.0%)** [Bound ≤ 8.81%] | 0/40 inventions observed in the period (consistent with a rate ≤ 8.81%); 3/40 answers omitted non-numeric nuances. |
| **3. `CLARIFY` and multi-turn census** | n = 30/30 (28 cat. + 2 OOC) | **29/30 (96.7%)** $\kappa_{HH}^{ord} = 0.91$, $AC_1 = 0.96$ | **27/28 (96.4% cat.)** 2/2 (100% OOC) | — | **0/30 (0.0%)** [Bound ≤ 11.57%] | **Drop-off of 53.33% (16/30)** with no turn-2 reply; continuation of **46.67% (14/30)**, with conditional convergence of **100% (14/14)** among those who replied. |
| **4. `ABSTAIN` census (`sin-contexto`)** | n = 48/48 (100% of OOC) | **47/48 (97.9%)** $\kappa_{HH}^{ord} = 0.90$, $AC_1 = 0.97$ | **44/48 (91.7%)** [80.0%, 97.7%] | — | **0/48 (0.0%)** [Bound ≤ 7.40%] | 91.7% were federal or municipal procedures; 4/48 (8.3%) were false abstentions caused by extreme colloquial jargon. |

*Note.* Because the share of positive cases exceeds 90% and the samples contain only 1 to 4 negative cases, Gwet's $AC_1$ tends toward high values close to raw agreement ($p_o$), while Cohen's κ is sensitive to marginal asymmetry. For transparency we report first the raw pre-adjudication agreement and the 2 × 2 matrices (a, b, c, d): Audit 1 (37, 1, 1, 1); Audit 2 (36, 1, 0, 3); Audit 3 (27, 0, 1, 2); and Audit 4 (43, 1, 0, 4). Brackets give exact 95% Clopper-Pearson intervals.

**Figure 4** shows the geometric partition of the N = 200 queries in the ($S_{cos}^{max}$, $S_{rr}^{(1)}$) plane of `clarify_policy`. **Figure 5** documents the **73× second-turn speedup** from attribute anchoring (`_FOLLOWUP_ATTR`: 0.42 s vs. 30.76 s with LLM rewriting) and the rise in phone coverage to **100.0%** through the L0 Directory.

**Figure 4**

*Empirical Distribution of the N = 200 Queries in the `clarify_policy` Decision Space*

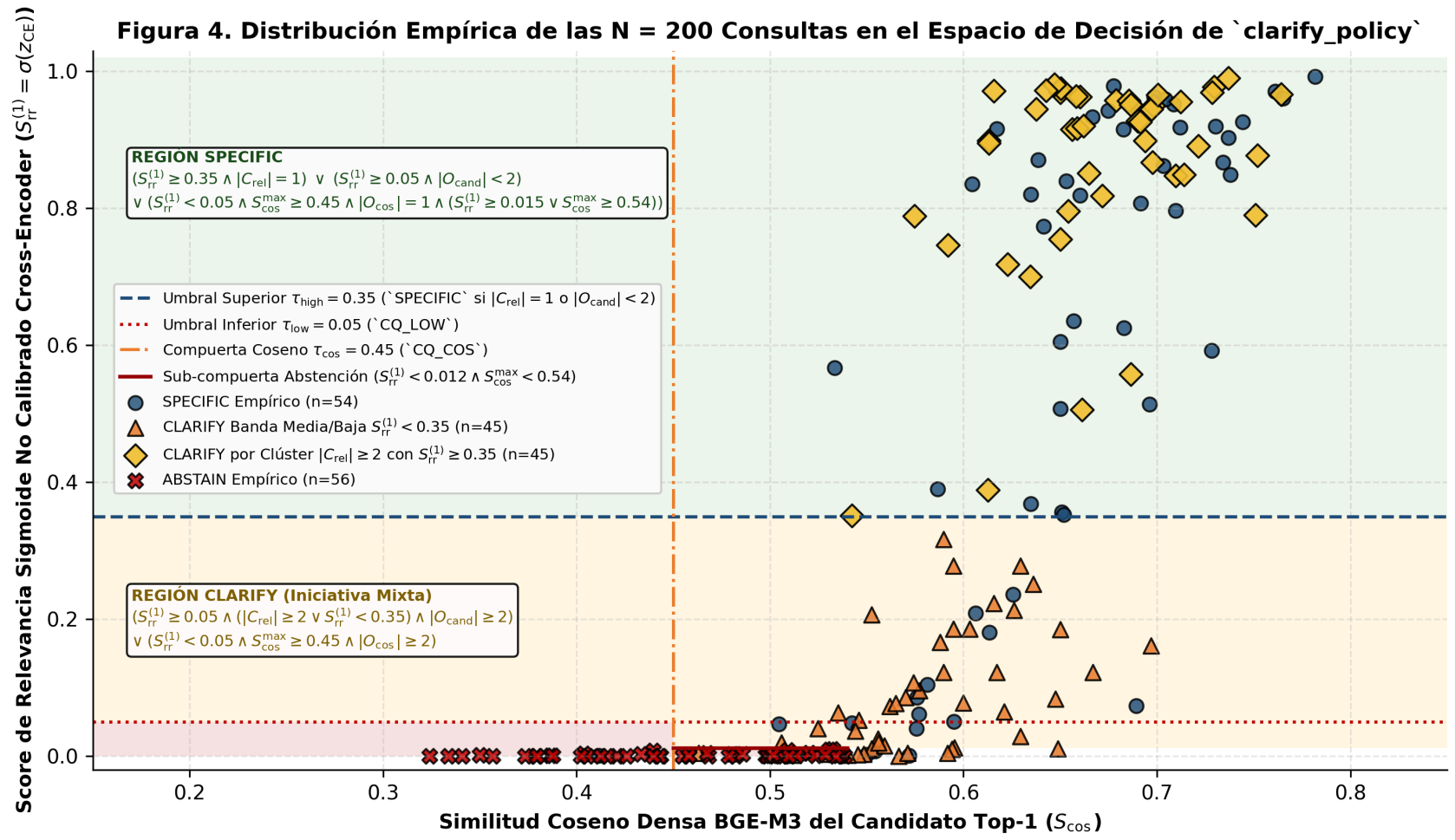

**Figure 5**

*Impact of Multi-Turn Anchoring, `CLARIFY` Convergence, and `AnyJev` Epistemic Separation*

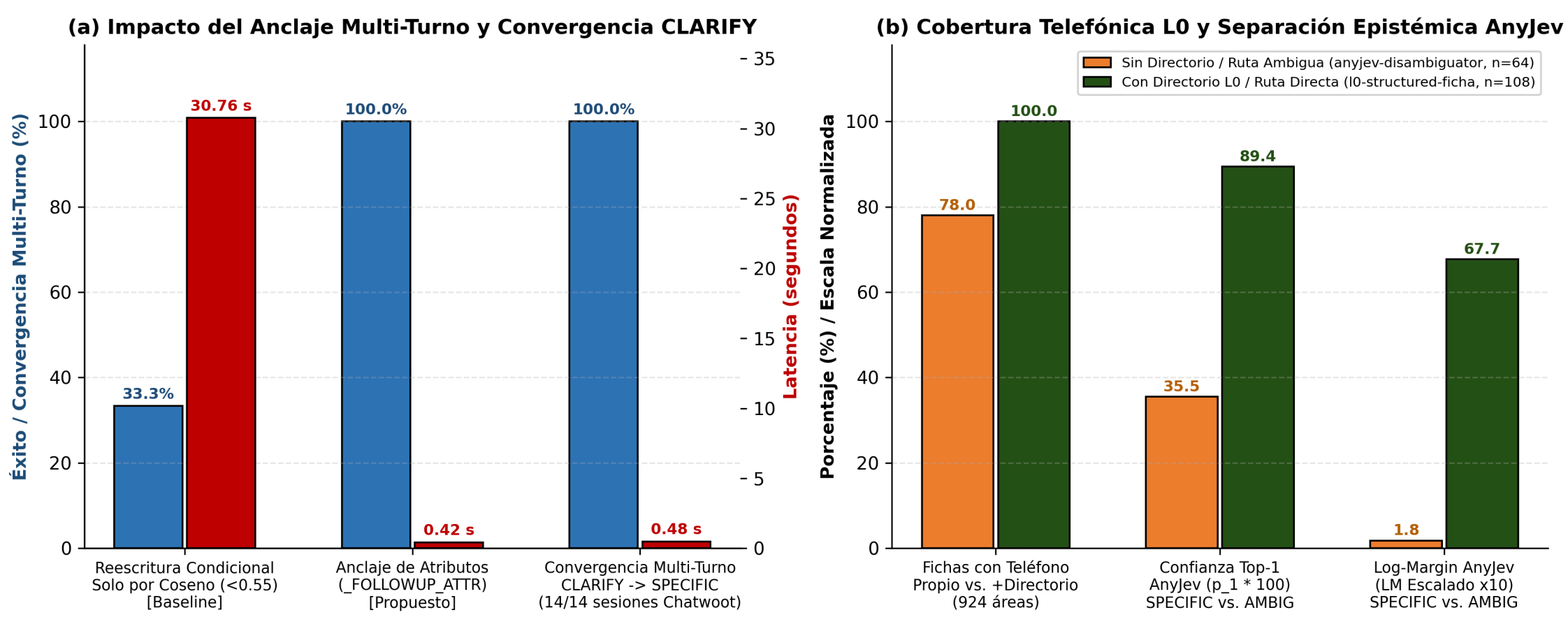

## 5. Discussion

### 5.1. Synthesis Against Research Questions ($RQ_0$–$RQ_3$)

The empirical results show that the technical and regulatory viability of a governed sovereign RAG system in subnational public administration ($RQ_0$) does not depend on deploying massive parameter models across dedicated GPU clusters (Branco et al., 2026), but on **governing uncertainty by inverting the generative pyramid**: resolving 89.47% of citizen traffic deterministically at Level 0 without free generation and restricting supervised synthesis with local LLMs (`3B−4B`) to the remaining 10.53%. This design explains the bimodal latency profile: direct Level 0 deliveries (68.95% of traffic) respond in sub-second median times ($p_{50}$ = 0.445 s; global median $p_{50}$ = 0.665 s), whereas routes that invoke CPU inference (`Query2Doc` prior to containment cutoffs or Level 1 synthesis) experience slow-tail queuing ($p_{95}$ = 30.412 s) when concurrent requests overlap.

Regarding the sub-questions: (i) for $RQ_1$, decoupling the procedural catalog from the phone directory prevents a -6.15 pp `Hit@1` loss from document brevity bias and preserves 93.33% `Hit@1` (100.0% `Hit@5`) across the strict long tail; (ii) for $RQ_2$, the tripartite gate outperforms the binary cutoff `B2` by +10.00 pp in strict Top-1 accuracy ($p_{Holm}$ = 0.01172) by replacing blind rejections with discriminating facet menus, though its real-world effectiveness is bounded by second-turn drop-off; and (iii) for $RQ_3$, the three-barrier numeric defense contained fee inventions across the generative census (0/40; population bound ≤ 8.81%) and intercepted all 40/40 synthetic perturbations.

### 5.2. Conversational Friction as the Operational Cost of Safe Prediction

A central methodological finding of this study is that safe coverage in mixed-initiative RAG carries an interaction cost invisible to static offline benchmarks. While the conditional upper bound reaches 93.00% under the assumption that every user replies to a `CLARIFY` menu, production telemetry reveals a **53.33% (16/30) drop-off rate** at Turn 2. Accounting for this retention factor ($\gamma$ = 0.4667) places the **dropout-adjusted effective resolution at 70.87%**. At digital government desks, guided clarification prevents regulatory misinformation, but introduces cognitive friction that requires reporting conditional safety bounds alongside dropout-adjusted effective resolution.

### 5.3. Non-Monotonic Trade-offs: Algorithmic Core (`C1`–`C5`) versus Domain Rules (`C6`–`C7`)

The seven-stage ablation (Tables 3A and 3B) highlights three architectural trade-offs:

1. **Separation between portable core and local rules:** The statistical-neural core (`C1`–`C5`) provides the transferable foundation (77.50% strict Top-1, 88.50% safe Top-5, and 63.43% effective resolution in `C5`), whereas deterministic domain rules (`C6`–`C7`) add +9.00 pp in strict Top-1 (86.50%), +4.50 pp in safe Top-5 (93.00%), and +7.44 pp in effective resolution (70.87%) by encoding portal-specific jurisdictional and welfare eligibility boundaries.
2. **Epistemic prudence in the long tail (`C3` → `C4`):** The shift in `Hit@1` from 100.0% to 93.33% upon activating the gate reflects deliberate containment: when procedures exhibit sister variants with narrow margins (LM < 0.35), the system requests clarification rather than risking an unconfirmed modality.
3. **Relevance inflation from generative expansion (`C4` → `C5` → `C6`):** Although `Query2Doc` raises in-catalog recall (`IN_HEAD` to 95.00%), it translates colloquial out-of-catalog complaints into plausible bureaucratic terminology, reducing strict abstention by -10.00 pp (68.57% → 58.57%). In public administration, generative query expansion must be constrained by explicit jurisdictional filters (`SCOPE_TERMS`).

## 6. Limitations and Threats to Validity

We synthesize seven methodological and operational limitations that bound the interpretation of our results:

1. **Four-tier scope of the numeric guardrail:** The verifier `verify_numeric_guardrail` ($N_{\$,UMA}(y) \subseteq N(d_{(1)})$) guarantees membership of generated numbers in the retrieved record (40/40 in functional tests; 0/40 inventions in Level 1, bound ≤ 8.81%). However, it does **not** prevent: (i) internal concept permutations within multi-fee records; (ii) Top-1 retrieval errors where an adjacent procedure is fetched (8.46% on the benchmark; 5.00% in Audit 1); or (iii) statutory reforms enacted after the latest catalog snapshot.
2. **Multi-turn drop-off penalty:** The 53.33% attrition rate following a `CLARIFY` menu lowers conditional safe coverage from 93.00% to an effective resolution of 70.87%. Despite 100% convergence among users who reply, this drop-off requires streamlining the selection UI.
3. **Benchmark scale (N = 200), pilot calibration ($N_{cal}$ = 60), and local rule overfitting (`C6`–`C7`):** While 5-fold stratified cross-validation (86.41% ± 7.50% strict Top-1) supports internal stability, initial thresholds derive from a pilot of $N_{cal}$ = 60 and rules `C6`–`C7` were crafted for this portal's lexical and jurisdictional patterns. Whereas the `C1`–`C5` core (77.50% Top-1) is directly portable, `C6`–`C7` carry an overfitting risk and require local recalibration and external validation in other jurisdictions.
4. **Limit against implicit sociodemographic bias:** The rule `strip_demographic_confounder` resolves leading self-introductions (5/5 pairs in `FAIRNESS_EXPLICIT`, functional check), but on socioeconomic traits embedded mid-sentence (`FAIRNESS_IMPLICIT`) pairwise invariance remains at **60.00% (6/10 pairs)**, confirming that syntactic rules do not replace counterfactually invariant representations (Yang et al., 2026).
5. **Homonymy across unindexed state branches (`OOC_UNINDEXED_STATE`):** For procedures from other state branches or autonomous bodies absent from the Executive catalog, 20.00% (4/20) receive a false `SPECIFIC` due to lexical overlap, accounting for four of the six false positives across the benchmark (6/70 = 8.57% in OOC; 6/200 = 3.00% overall).
6. **Divergence between median latency ($p_{50}$) and slow-tail CPU queuing ($p_{95}$):** The sub-second global median ($p_{50}$ = 0.665 s; $p_{50}$ = 0.445 s in direct L0) does not represent slow-route performance or behavior under concurrent load: when 2–3 overlapping requests trigger `Query2Doc` (including the 20.53% of L0 containment at $p_{50}$ = 6.410 s) or Level 1 synthesis ($p_{50}$ = 20.150 s), serialization across the 8 vCPUs pushes the global tail to $p_{95}$ = 30.412 s ($p_{99}$ = 35.626 s).

7. **Shallow regex anonymization and internal institutional auditing:** The `_sanitize_pii` filter masks structured identifiers (`CURP`, `RFC`, phone numbers, emails), but not free-text personal names. Moreover, double-blind audits were conducted by state public-sector analysts, warranting future evaluation with external citizen panels.

**Reproducibility, Software Environment, and Data Availability (`scripts_y_datos/`, v1.0.0-R3, `MIT` and `CC-BY 4.0` licenses):** The execution environment uses `Python 3.11+`, `PostgreSQL 16` with `pgvector 0.7.0`, `BAAI/bge-m3`, `BAAI/bge-reranker-v2-m3`, `anyjev==0.3.0`, and `Ollama 0.5.x` (`qwen2.5:3b-instruct-q4_K_M` and `qwen3:4b`). All open artifacts are included in the supplementary package (**Technical Supplement, Tables S1 and S5**) and can be deterministically verified in < 2 s by running `python3 scripts_y_datos/verify_all_paper_claims.py`: (i) `ablation_study_200_r3.json` (N = 200 × 9 configurations: `B1`, `B2`, `C1`–`C7`); (ii) `production_audit_metrics_r3.json` and `guardrail_probe_results.json` (n = 40 clean and n = 40 with controlled numeric perturbation); and (iii) `anyjev_decider.py`, `directorio_helper.py`, `directorio_consolidado.json` (924 areas), `run_ablation_200_r3.py`, `run_external_baselines.py`, `run_guardrail_probe.py`, and `README_REPLICACION_v1.0.0-R3.md`. Under the state privacy statute (LPDPPSOET), raw production conversation logs (N = 380) are not released in open text; their aggregated matrices and audited epistemic statistics are provided instead.

## 7. Conclusions

This study showed that a subnational public administration can operate a governed sovereign RAG system on local infrastructure (`CPU + NPU`) by inverting the generative pyramid: in production (N = 380), **89.47%** of traffic is resolved through deterministic Level 0 routes (**68.95%** direct deliveries and **20.53%** epistemic containment) and only **10.53%** escalates to supervised generative synthesis at Level 1. Joint evaluation in production and on the stratified benchmark (N = 200) supports four main conclusions:

1. **First-turn performance, conditional bound, and conversational friction:** The full system (`C7`) resolves **51.50%** of queries immediately in the first turn and achieves **86.50%** strict Top-1 accuracy (+10.00 pp over the binary classifier `B2`, p = 0.00453, $p_{Holm}$ = 0.01172). While the **conditional safe Top-5 upper bound reaches 93.00%**, discounting the **53.33%** citizen drop-off observed after a `CLARIFY` menu places **dropout-adjusted effective resolution at 70.87%**, confirming that interactive safety metrics must always be paired with their attrition cost.
2. **Differentiated external validity between portable core (`C1–C5`) and local rules (`C6–C7`):** The statistical-neural core (`C1`–`C5`) achieves **77.50%** strict Top-1 accuracy and **88.50%** safe Top-5 coverage (63.43% effective resolution) on its own, constituting the directly transferable component across administrations. Deterministic domain rules (`C6`–`C7`) add +9.00 pp in strict Top-1 and +4.50 pp in safe Top-5, but because they encode portal-specific patterns over N = 200 queries, they carry an overfitting risk and require local recalibration in other jurisdictions.
3. **Long-tail retrieval without popularity bias:** Decoupling the catalog (1,143 procedures) from the directory (924 areas), combined with bounded structural boosts, reaches **91.54%** `Hit@1` and **93.85%** `Hit@5` in catalog, preserving **93.33%** `Hit@1` and **100.0%** `Hit@5` on strict long-tail procedures with no historical visits.
4. **Strict delimitation of numeric safety and operational latency:** Across the Level 1 generative census no fee inventions were observed (0/40, population upper bound ≤ 8.81% at 95% confidence; 40/40 synthetic injections intercepted), with residual errors concentrated in adjacent-procedure retrieval (5.00% in production) and out-of-catalog false `SPECIFIC` decisions (6/70 = 8.57% in OOC; 3.00% of the benchmark). Likewise, habitual median speed ($p_{50}$ = 0.665 s global; $p_{50}$ = 0.445 s in direct L0) must be strictly distinguished from slow-tail latency ($p_{95}$ = 30.412 s, $p_{99}$ = 35.626 s) caused by CPU queuing of `Query2Doc` and Level 1 requests.

Future work will prioritize four directions derived from these limitations: (i) executing the jurisdictional filter before `Query2Doc` to cut abstention latency to 0.38 s and reduce the CPU $p_{95}$ tail; (ii) adding inter-branch referral records to mitigate false `SPECIFIC` answers from institutional homonymy; (iii) adopting counterfactually invariant representations against implicit socioeconomic bias (60.00% in `FAIRNESS_IMPLICIT`); and (iv) conducting external validations with independent citizen panels and catalogs from other government tiers.

### Declaration of Generative AI and AI-Assisted Technologies in the Writing Process

During the preparation of this work the author(s) used large language model-based editing and translation assistance (Claude / Gemini) in order to improve the readability, English language translation, and formatting of the manuscript. After using this tool/service, the author(s) reviewed and edited the content as needed and take(s) full responsibility for the content of the publication.

## 8. References

Aliannejadi, M., Zamani, H., Crestani, F., & Croft, W. B. (2019). Asking clarifying questions in open-domain information-seeking conversations. In *Proceedings of the 42nd International ACM SIGIR Conference on Research and Development in Information Retrieval (SIGIR '19)* (pp. 475–484). Association for Computing Machinery. https://doi.org/10.1145/3331184.3331265

Alqahtani, A., Yan, J., Vlahu-Gjorgievska, E., & Cao, C. (2026). Factors influencing citizens' adoption of AI-enabled public services in Saudi Arabia: The role of cybersecurity, privacy, and trust. *ACM Digital Government: Research and Practice*, *7*(2), Article 14. https://doi.org/10.1145/3785153

Arguelles Toache, E., & Amaro Rosales, M. (2023). Preocupaciones éticas en el uso de inteligencia artificial, transparencia y derecho de acceso a la información. El caso de los chatbots en el gobierno de México, en el contexto de la COVID-19 [Ethical concerns in the use of artificial intelligence, transparency, and the right of access to information: The case of chatbots in the Mexican government during COVID-19]. *Estudios en Derecho a la Información*, (15), 65–91. https://doi.org/10.22201/iij.25940082e.2023.15.17472

Branco, A., Gomes, L., Santos, R., Santos, E., Silva, J., Marques, N., & Rodrigues, M. (2026). Sovereign AI-based public services are viable and affordable. In *Proceedings of the Joint International Conference on Computational Linguistics, Language Resources and Evaluation (LREC 2026)*. https://doi.org/10.48550/arXiv.2603.01869

Camino-Moreno, S. (2026). A RAG-driven conversational system for citizen information requests at Ecuador's tax authority. *Journal of Emerging & Sustainable Smart Engineering (JESSE)*, *2*(1), 1–13. https://doi.org/10.31243/jesse.uta.v2i1.2998.2026

Cejudo, G. M., & Michel, C. L. (2017). Addressing fragmented government action: Coordination, coherence, and integration. *Policy Sciences*, *50*(4), 745–767. https://doi.org/10.1007/s11077-017-9281-5

Chen, J., Lin, H., Han, X., & Sun, L. (2024). Benchmarking large language models in retrieval-augmented generation. In *Proceedings of the AAAI Conference on Artificial Intelligence* (Vol. 38, No. 16, pp. 17754–17762). https://doi.org/10.1609/aaai.v38i16.29728

Cheng, M., Luo, Y., Ouyang, J., Liu, Q., Liu, H., Li, L., Yu, S., Zhang, B., & Chen, E. (2026). A survey on knowledge-oriented retrieval-augmented generation. *ACM Transactions on Information Systems*, *44*(8), Article 170. https://doi.org/10.1145/3833415

Cormack, G. V., Clarke, C. L. A., & Buettcher, S. (2009). Reciprocal rank fusion outperforms condorcet and individual rank learning methods. In *Proceedings of the 32nd International ACM SIGIR Conference on Research and Development in Information Retrieval (SIGIR '09)* (pp. 758–759). Association for Computing Machinery. https://doi.org/10.1145/1571941.1572114

Criado, J. I., Sandoval-Almazán, R., & Gil-Garcia, J. R. (2025). Artificial intelligence and public administration: Understanding actors, governance, and policy from micro, meso, and macro perspectives. *Public Policy and Administration*, *40*(1), 3–18. https://doi.org/10.1177/09520767241272921

Cruz-Meléndez, C., Jiménez Canseco, J. M., & Martínez Gutiérrez, R. (2024). La implementación de chatbots en gobiernos municipales mexicanos. Exploración y diagnóstico de uso [The implementation of chatbots in Mexican municipal governments: Exploration and usage diagnosis]. *Revista Mexicana de Análisis Político y Administración Pública (REMAP)*, *13*(26), 119–153. https://doi.org/10.15174/remap.v13i26.441

de Oliveira, H. T. A., César Sobrinho, Á. A. D. C., da Silva, L. D., & Perkusich, A. (2026). Large language models for chatbot applications handling sensitive information. *Expert Systems With Applications*, *299*, 130145. https://doi.org/10.1016/j.eswa.2025.130145

Farquhar, S., Kossen, J., Kuhn, L., & Gal, Y. (2024). Detecting hallucinations in large language models using semantic entropy. *Nature*, *630*, 625–630. https://doi.org/10.1038/s41586-024-07421-0

Feng, S., Shi, W., Wang, Y., Ding, W., Balachandran, V., & Tsvetkov, Y. (2024). Don't hallucinate, abstain: Identifying LLM knowledge gaps via multi-LLM collaboration. In *Proceedings of the 62nd Annual Meeting of the Association for Computational Linguistics (ACL 2024, Vol. 1: Long Papers)* (pp. 14664–14690). Association for Computational Linguistics. https://doi.org/10.18653/v1/2024.acl-long.786

Friel, R., Belyi, M., & Sanyal, A. (2024). *RAGBench: Explainable benchmark for retrieval-augmented generation systems* (Preprint arXiv:2407.11005). arXiv. https://doi.org/10.48550/arXiv.2407.11005

Galicia Hernández, F. M., Gómez Álvarez, C., Heredia Sánchez, I., Rojas Lita, B., Peeters, R., & Rentería, C. (2025). Aprender de primera mano: ¿La experiencia ciudadana de cargas administrativas afecta la confianza en el gobierno y la percepción sobre su desempeño? [Learning first-hand: Does citizens' experience of administrative burden affect trust in government and perceptions of its performance?]. *Gestión y Política Pública*, *34*(2), 1–32. https://doi.org/10.60583/gypp.v34i2.8478

Goyal, A., Mukherjee, K., Saxena, A., Phukan, A., Chandrasekharan, E., & Sundaram, H. (2026). Masking or mitigating? Deconstructing the impact of query rewriting on retriever biases in RAG. In *Findings of the Association for Computational Linguistics: ACL 2026* (pp. 8517–8530). Association for Computational Linguistics. https://aclanthology.org/2026.findings-acl.414/

Joren, H., Zhang, J., Ferng, C.-C., Juan, D.-C., Taly, A., & Rashtchian, C. (2025). Sufficient context: A new lens on retrieval-augmented generation systems. In *Proceedings of the International Conference on Learning Representations (ICLR 2025)*. https://doi.org/10.48550/arXiv.2411.06037

Kamalipour, A. A., Asadi, S., & Amiri Chimeh, M. M. (2026). From vectors to knowledge graphs: A comprehensive analysis of modern retrieval-augmented generation architectures. *Computer Science Review*, *61*, 100925. https://doi.org/10.1016/j.cosrev.2026.100925

Kamath, A., Jia, R., & Liang, P. (2020). Selective question answering under domain shift. In *Proceedings of the 58th Annual Meeting of the Association for Computational Linguistics (ACL 2020)* (pp. 5684–5696). Association for Computational Linguistics. https://doi.org/10.18653/v1/2020.acl-main.503

La, D.-D., & Tran, K. (2024). Analysis of document retrieval for online public administrative procedure services. *Journal of Science and Technology on Information and Communications*, *2024*(2), 54–63. https://doi.org/10.32913/mic-ict-research-vn.v2024.n2.1297

Lei, J., Kun, W., Chen, Z., & Xin, Z. (2026). NaviRAG: Learning to navigate knowledge graphs for retrieval-augmented generation. In *Proceedings of the ACM SIGKDD Conference on Knowledge Discovery and Data Mining (KDD '26)* (pp. 2521–2532). Association for Computing Machinery. https://doi.org/10.1145/3770855.3817935

Li, R., Cao, S., Liu, R., & Duprey, A. (2026). EviDex: Provenance-weighted evidence-path indexing for fresh and auditable retrieval under continuous updates. *Proceedings of the ACM on Measurement and Analysis of Computing Systems (SIGMETRICS)*, *10*(2), Article 38. https://doi.org/10.1145/3805636

Liu, B., Hu, Y., Ai, Q., Liu, Y., Wu, Y., Li, C., & Shen, W. (2025). Generating clarifying questions for conversational legal case retrieval without external knowledge. *ACM Transactions on Information Systems*, *43*(4), Article 102. https://doi.org/10.1145/3736161

Lupart, S., Mekonnen, K. A., Abbasiantaeb, Z., & Aliannejadi, M. (2026). uva-irlab-conv at SemEval-2026 Task 8: Multi-turn RAG with learned sparse retrieval and listwise reranking. In *Proceedings of the 20th International Workshop on Semantic Evaluation (SemEval-2026)* (pp. 2972–2980). Association for Computational Linguistics. https://aclanthology.org/2026.semeval-1.373/

Mala, C. S., Di Maio, C., Proietti, M., Gezici, G., Giannotti, F., Melacci, S., Lenci, A., & Gori, M. (2025). Towards building a trustworthy RAG-based chatbot for the Italian public administration. *Frontiers in Artificial Intelligence and Applications*, *406*, 215–222. https://doi.org/10.3233/FAIA250637

Malekzadeh, M. (2026). Bringing back the front desk: A bounded case for AI-first public services. *ACM Digital Government: Research and Practice*, *7*(3), Article 17. https://doi.org/10.1145/3821442

Moynihan, D., Herd, P., & Harvey, H. (2015). Administrative burden: Learning, psychological, and compliance costs in citizen-state interactions. *Journal of Public Administration Research and Theory*, *25*(1), 43–69. https://doi.org/10.1093/jopart/muu009

Neguse, M., Gabay, D., Reychav, I., McHaney, R., & Jonathan, G. M. (2026). Citizen sentiment and public value in AI-mediated public services: A case study of a ministry chatbot. In *Proceedings of the ACM SIGMIS Computers and People Research Conference (SIGMIS-CPR '26)*. Association for Computing Machinery. https://doi.org/10.1145/3768310.3807801

Niu, M., Su, J., & Tian, C. (2026). UA-RAG: Uncertainty-aware dynamic retrieval-augmented generation. *Neurocomputing*, *699*, 134357. https://doi.org/10.1016/j.neucom.2026.134357

Phillips, E., Gustafsson, F. K., Wu, S., Thakur, A., & Clifton, D. A. (2026). *Entropy alone is insufficient for safe selective prediction in LLMs* (Preprint arXiv:2603.21172). arXiv. https://doi.org/10.48550/arXiv.2603.21172

Prencipe, G., Tommasi, A., Zavattari, C., Tesi, G., Storchi, L., & Shahin, K. (2026). Transforming public administration workflows with multi-agent AI: A human-in-the-loop knowledge orchestration framework (CARE). *Computers in Human Behavior Reports*, *23*, 101253. https://doi.org/10.1016/j.chbr.2026.101253

Ramírez-Hernández, P., Valle-Cruz, D., & Mendoza-Méndez, R. V. (2023). Revisión de los chatbots basados en inteligencia artificial en la administración pública: Hacia una arquitectura para el gobierno [A review of AI-based chatbots in public administration: Toward an architecture for government]. *Espacios Públicos*, *24*(60), e21317. https://doi.org/10.36677/espaciospublicos.v23i60.21317

Rappan, J., Blasi, M., Blanco, M. C., Micchele, L., Galaviz, D., Balp, E., Albuquerque, H., Kulowski, L., & Fernandes, R. (2025, August 28). Meet Boti: The AI assistant transforming how the citizens of Buenos Aires access government information with Amazon Bedrock [Architecture technical report]. *AWS Machine Learning Blog*. https://aws.amazon.com/blogs/machine-learning/meet-boti-the-ai-assistant-transforming-how-the-citizens-of-buenos-aires-access-government-information-with-amazon-bedrock/

Raza, S., Sapkota, R., Karkee, M., & Emmanouilidis, C. (2026). TRiSM for agentic AI: A review of trust, risk, and security management in LLM-based agentic multi-agent systems. *AI Open*, *7*, 71–95. https://doi.org/10.1016/j.aiopen.2026.02.001

Saad-Falcon, J., Khattab, O., Potts, C., & Zaharia, M. (2024). ARES: An automated evaluation framework for retrieval-augmented generation systems. In *Proceedings of the 2024 Conference of the North American Chapter of the Association for Computational Linguistics (NAACL 2024)* (pp. 338–354). Association for Computational Linguistics. https://doi.org/10.18653/v1/2024.naacl-long.20

Sánchez-Navalón, H., Monserrat, C., Garigliotti, D., & Ferri, C. (2024). Evaluating performance and trustworthiness of RAG systems for generating administrative text. In *Intelligent Data Engineering and Automated Learning – IDEAL 2024* (Lecture Notes in Computer Science, Vol. 15245, pp. 410–421). Springer. https://doi.org/10.1007/978-3-031-77731-8_37

Sekulić, I., Aliannejadi, M., & Crestani, F. (2021). User engagement prediction for clarification in search. In *Advances in Information Retrieval (ECIR 2021)* (Lecture Notes in Computer Science, Vol. 12656, pp. 619–633). Springer. https://doi.org/10.1007/978-3-030-72113-8_41

Siro, C., Yuan, Y., Aliannejadi, M., & de Rijke, M. (2026). AGENT-CQ: Automatic generation and evaluation of clarifying questions for conversational search with large language models. *ACM Transactions on Information Systems*, *44*(5), Article 110. https://doi.org/10.1145/3809182

Song, M. (2026). Defining the problem: The impact of OCR quality on retrieval-augmented generation performance and strategies for improvement. *Information Processing and Management*, *63*, 104368. https://doi.org/10.1016/j.ipm.2025.104368

Tanjim, M. M., In, Y., Chen, X., Bursztyn, V. S., Rossi, R. A., & Dernoncourt, F. (2025). Disambiguation in conversational question answering in the era of LLMs and agents: A survey. In *Proceedings of the 2025 Conference on Empirical Methods in Natural Language Processing (EMNLP 2025)* (pp. 9537–9550). Association for Computational Linguistics. https://doi.org/10.18653/v1/2025.emnlp-main.482

Wang, L., Yang, N., & Wei, F. (2023). Query2doc: Query expansion with large language models. In *Proceedings of the 2023 Conference on Empirical Methods in Natural Language Processing (EMNLP 2023)* (pp. 9414–9423). Association for Computational Linguistics. https://doi.org/10.18653/v1/2023.emnlp-main.585

Wigger, L., Podolsky, D., Wilmink, T., & Peng, Y. (2026). RAGTUM at SemEval-2026 Task 8: Contextual query rewriting and dense retrieval for multi-turn RAG. In *Proceedings of the 20th International Workshop on Semantic Evaluation (SemEval-2026)*. Association for Computational Linguistics. https://aclanthology.org/2026.semeval-1.227/

Xu, J., Li, X., Song, Q., & Zhu, S. (2026). MoRE-RAG: Bayesian optimal fusion of multiple retrieval experts for retrieval-augmented generation. *Neurocomputing*, *701*, 134536. https://doi.org/10.1016/j.neucom.2026.134536

Yang, B., Yu, X.-H., Zheng, X., Nong, J., Liu, Z., Dai, X.-M., & Xie, X. (2025). CuBE: A customizable bounds evaluation framework for automated assessment of RAG systems in government services. *Applied Sciences*, *15*(19), 10447. https://doi.org/10.3390/app151910447

Yang, Y., Xu, Z., Li, J., Ma, C., Chan, J., Sanderson, M., Zheng, X., & Ren, Y. (2026). Mitigating bias in large language model based question answering through causal front door prompting. In *Proceedings of the ACM SIGIR Conference on Research and Development in Information Retrieval (SIGIR '26)*. Association for Computing Machinery. https://doi.org/10.1145/3805712.3809679

Ye, X., Xu, C., Zhang, C., Du, Z., Xu, J., Wang, G., & Dong, Z. (2025). Q-PRM: Adaptive query rewriting for retrieval-augmented generation via step-level process supervision. In *Findings of the Association for Computational Linguistics: EMNLP 2025* (pp. 15210–15224). Association for Computational Linguistics. https://doi.org/10.18653/v1/2025.findings-emnlp.817

Zamani, H., Dumais, S. T., Craswell, N., Bennett, P. N., & Lueck, G. (2020). Generating clarifying questions for information retrieval. In *Proceedings of The Web Conference 2020 (WWW '20)* (pp. 418–428). Association for Computing Machinery. https://doi.org/10.1145/3366423.3380126

Zhang, J., Yang, T., Shi, Y., Chen, E., Nie, Z., Wan, K., Hong, L., Liu, N., & Wu, L. (2026). *AnyJev: Typed decisions from one prefill of any open LLM* (Preprint arXiv:2610.00831v1; software v0.3.0). Nokia Applied Research, Tencent Hunyuan & The Hong Kong Polytechnic University. https://pypi.org/project/anyjev/